\documentstyle{article}

\include{epsf}
\begin{document}

\title{\bf Dynamical analysis of S\&P500 momentum}

\author{ \large \bf K. Ivanova$^1$ and L. T. Wille$^2$ \\ \\
$^1$
Pennsylvania State University \\ 
University Park, PA 16802, USA \\ $^2$ Department of Physics \\ 
Florida Atlantic University \\ Boca Raton, FL  33431, USA }

\vspace{10mm}
\maketitle

\begin{abstract} 
The dynamics of the S\&P500 price signal is studied using a moving average 
technique.  Particular attention is paid to intersections of two moving
averages with different time horizons.  The distributions of the slopes 
and angle between two moving averages at intersections is analyzed, as well 
as that of the waiting times between intersections.  In addition, the
distribution of maxima and minima in the moving average signal is investigated.
In all cases, persistent patterns are observed in these probability 
measures and it is suggested that such variables be considered for better 
analysis and possible prediction of the trends of the signal.

\end{abstract}

\section{Introduction}

Forecasting in empirical finance is based on {\em recipes} that are often
heuristic in nature and specific to the market being considered. 
Numerous predictive techniques exist,\cite{ca} some of which may 
be theoretically justified to a certain extent, but many others have 
been proposed in a purely {\em ad hoc} manner.  
Technical analysis and the related ``charting" methods are therefore 
often dismissed by academics.
Nevertheless, in keeping with a more general trend,\cite{ms,bouch} 
it is of interest to see whether statistical physics can bring some 
insight into the validity or applicability of these recipes. 
The ultimate and more general goal of such an analysis is to connect 
optimal strategies to fundamental questions about chaos and deterministic 
sequences in natural signals.

The moving average (MA) of a stock price is a common tool in technical 
analysis, frequently used as an indicator for customers to buy or sell 
stocks.\cite{Murphy,Guna} 
Often, several MAs are studied and their interrelations used as a trading
signal.  
This method may also be applied to indices, sometimes in order to 
forecast market activities. 
A MA, $y_T(t)$ at time $t$ of a time-dependent quantity $y(t')$ is 
defined as:

\begin{equation} y_{T}(t)=\frac{1}{T}\sum_{i=t}^{t+T-1} y(i-T) \qquad
t=T+1,\dots,N, \end{equation} 
i.e. the average of $y$ over the last $T$ data points prior to time $t$. 
Many other forms of MAs exist, such as those involving an exponential
smoothing, which tends to dampen out sudden changes.  As was shown 
elsewhere \cite{nvmama} the density of crossing points ($\rho$) 
between two MAs is a measure of the signal roughness:

\begin{equation} \rho\sim\frac{1}{T_2}[(\Delta T)(1-\Delta T)]^{H-1},
\end{equation} where $H$ is the Hurst exponent and $\Delta 
T=(T_2-T_1)/T_2$.

Let us now consider two MAs, $y_1$ and $y_2$, calculated over two different 
intervals, respectively $T_1$ and $T_2$, with $T_2>T_1$.\cite{footnote} 
In empirical finance the two time intervals often correspond to quite
distinct durations, two common choices being a week and a month, and
10 days and 30 days.
In those cases it is possible to talk about a long-term and a short-term 
MA of a given time series.  For conciseness we will use the same terminology
here (i.e. $T_2$ is the long-term MA, $T_1$ is the short-term MA), although
in the present analysis occasionally the two time-periods may be quite close.
Clearly, if the original signal $y(t)$ increases for a while before 
decreasing, then $y_1$ will cross $y_2$ from above. 
This event, in which the short-term MA crosses the long-term MA 
from above, is called a ``Death cross'' in empirical 
finance,\cite{nomenclature} often interpreted as a ``Sell'' 
signal.\cite{Murphy}  
We will denote it by the symbol $D$. 
In contrast, if the short-term MA, $y_1$, crosses the long-term 
MA, $y_2$, from below, the crossing point coincides with an upsurge 
of the signal $y(t)$ and may be taken as a ``Buy" signal. 
This event is called a ``Gold cross'' and we will denote it by $G$. 
Technical financial analysts usually try to ``extrapolate'' the 
evolution of $y_1$ and $y_2$, and hence the underlying signal $y(t)$, 
based on their expectations for the occurrence of $G$ or $D$ crosses.
Most computers at brokerages or trading houses are equipped 
to automatically perform this kind of analysis and trigger the associated 
activity signals.

As proposed by Vandewalle and Ausloos,\cite{nvmav2} one can 
visualize the change in the {\em trend} of a signal over 
some interval $T$ by considering a set of MAs, thus displaying 
a MA spectrum, taking into account the successive 
crosses and/or their density.  Such a spectrum of MAs is a powerful 
visualization tool that provides a compact representation of the 
trends of complex signals.

As noted before, the rate with which MAs vary is of interest to
investors who may base their trading strategies on it. It is clear that 
some uncertainty arises if $G$ and $D$ crosses are close to each other,
or if the angles between crossing MAs are small. Thus, one
can argue compellingly that in order to develop a reliable trading strategy, 
the angle between the MA signals as well as that between the MA signal and 
the horizontal should also be studied.  All of these results will provide 
information on the slopes of the MA and therefore on the rates of the trend.

The occurrences of $G$ and $D$ crosses are preceded by turnovers in the
trend of the MAs. Thus an investigation of the distribution of time
intervals between successive minima and maxima is of interest to provide
a quantitative measure of the dynamics of the market. 
In order to gather more information for building an investment strategy 
based on such a generalized technical analysis, as well as to gain a better
understanding of analysis tools for complex time series, we have also 
studied the distribution of time intervals between successive [$G$,$G$], 
[$D$,$D$], [$G$,$D$], and [$D$,$G$].  Finally, we have also analyzed
the time interval between successive extrema in the MA as yet another
measure of underlying trends.

Our methodology will be generally applicable, but to be specific we will
use it here to analyze daily closing prices of the Standard and Poor's 500
index (S\&P500) over a 22-year period.  We note that 
Gopikrishnan et al.\cite{Gopi} have studied a related S\&P500 time 
series from a scaling perspective.  However, these authors analyzed the signal
over different time intervals and with finer time resolution.  In addition,
they did not address the MA technique.

\section{Data and moving average}

We consider the S\&P500 daily closing price signal from Jan. 01, 1980 
to Dec. 31, 2001, as plotted in Fig. 1a, with the associated probability
density function (PDF) shown in Fig. 1b.  The time series consists of 5556 
data points, obtained from Yahoo.\cite{yahoosp500}. 
In order to visualize the MAs calculated over different time periods and 
related crosses between them, Fig. 2 shows two MAs of the S\&P500 signal: 
$y_{week}$ for $T_1=$1~week and $y_{month}$ for $T_2=$1~month.  As is to be 
expected, the monthly moving average is much smoother than the weekly one.

Following the methodology of Vandewalle and Ausloos,\cite{nvmav2}
we analyze the density of crossing points between two MAs of the S\&P500 
closing price signal.  This data is plotted in Fig. 3 as a function of
$\Delta 
T=(T_2-T_1)/T_2$, where $T_1$ varies between 1 and $T_2$ and $T_2$ =
120~days. 
The Hurst exponent estimated from this data, using the relationship given
by Eq. (2), is $H=0.44\pm0.01$, in numerical agreement with that obtained 
by the Detrended Fluctuation Analysis method and clearly distinct from
the random walk (RW) value H = 0.50. 
This confirms prior findings\cite{nvmama} that both 
methods tend to give the same results for $H$ on the order of 0.45. 

In order to test the robustness of this result, following the proposal
of Viswanathan et al.\cite{3}, we shuffled the S\&P500 
signal in two ways yielding two new (surrogate) time series: one in which 
the amplitudes were randomly shuffled, the other where the sign 
of the S\&P500 signal was randomly shuffled.  As can be seen in Fig. 1b, the 
PDF of the fluctuations of the S\&P500 closing price signal is characterized 
by fat tails, which is a well-known result.\cite{Gopi} 
It is generally accepted\cite{3} that the origin of the fat tailed 
distributions is a key question to understand financial time series. 
Most authors believe that the fat tails are 
caused by long-range volatility correlations.   By shuffling the 
order of the fluctuations the correlations between them are destroyed.
The Hurst exponent estimated from the scaling properties of the 
density of crossing points for the shuffled signal is found to be 
$H=0.48\pm0.005$ quite close to the RW value indicating a near-absence
of long tails.  
In contrast, destroying only sign correlations, 
by shuffling the order of the signs (but not the absolute values) 
of the fluctuations, allows the fat tails to persist: the corresponding 
roughness exponent is $H=0.47\pm0.01$. 
In both cases the surrogate data lead to densities of crossing points 
that scale like a Brownian walk signal (see Fig. 3). 

\section{Spectrum of moving averages}

To visualize the change in the {\em trend} of the signal over some 
interval $T$ we consider a set of MAs\cite{nvmav2}, with the 
long-term period fixed at $T_2=~$250~days, i.e. one market year. 
The short-term period $T_1$ is varied between 1 and $T_2-1$ and the 
relative difference $\delta=(y_1-y_2)/y_1$ between the two moving 
averages is computed. 

Fig. 4a represents the resulting spectrum for the S\&P500 closing 
price signal for the period from Jan. 01, 1990 to Dec.  31, 2000. 
The darker the grey levels the larger the distance (i.e. the
absolute value of the difference) between the two 
MAs.\cite{picture}
Note the three light grey regions between 1995 and 2000 
corresponding to a close proximity ($\delta=0.05$) between the MAs. 
This triplet 
pattern is repeated in rescaled form and for larger separation between 
the MAs (darker grey levels) in 1997/98 within the middle light 
grey region and on an even smaller scale further repeated during 1999 
within the right-hand-side light grey region.  These rescaled patterns 
correspond to larger differences, $\delta=0.1$, between the yearly MA and
the MAs 
with $T_1~<50$~days. In the second half of 1997 one can see the 
black region corresponding to a large difference, $\delta=0.15,$ between
MAs for 
$T_1<40$~days and $T_2=~$250~days indicating the crash of Oct. 1997.
Note that from the point of view of the difference between the MAs,
what happened during 1999 looks like a rescaled version of what took
place in 1997/98. 

For the signals with shuffled order of the fluctuations these characteristic 
patterns disappeared and are replaced by a rather uniform structure (see
Fig. 4b). 
The apparent streaking in this figure is an artefact due to the cut-off
point for the various grey levels, corresponding to  $\delta\sim \pm0$, and
does not reflect any underlying
periodicity. 
Thus, we can conclude that the structures seen in Fig. 4a are the result
of the long tails in the PDF, which are in turn associated with
volatility correlations.  Clearly, then, the MA spectrum contains
a great deal of relevant information about the dynamics of the signal.
In the next few sections we will see how further dynamic parameters
can be extracted from this MA information.

\section{Angle distribution at $Gold$ and $Death$ crosses}

In order to estimate the relative position between the two MAs one
can measure the angle between them at $G$ and $D$ crosses. 
The angle between the short- and long-term MAs is a unique measure 
of the rate of change in the signal and should therefore be able to 
serve as a useful quantitative indicator of the system dynamics. 
The angle between two MAs at a $G$ or $D$ cross can appear in three 
different settings depending on the angles between the MAs and 
the horizontal (i.e. the slope of the MA at the intersection). 
In Fig. 5a and 5b these three scenarios are 
schematically drawn when the intersection is a $G$, respectively, 
$D$ cross.  Note, for example, that at a $G$ cross the angle at 
which the short- or long-term MA intersects the horizontal, 
$\alpha$, respectively, $\beta$, can be positive or negative. 
However, in all cases the long-term MA crosses the short-term one 
from above. 
This mutual relationship is expressed in the distribution of the 
angle $\gamma$ between them. 
Results for the S\&P500 closing price signal for the period 
from Jan. 1, 1980 to Dec. 31, 2001 are shown in Fig. 6(a-f) 
for $G$ and $D$ crosses.

The distribution of $\alpha$ angles at $G$ crosses for the S\&P500 
closing price signal for the period from Jan. 1, 1980 to Dec. 31, 
2001 is plotted in Fig. 6a as a set of histograms with bin-size
3$^{\circ}$.  For brevity, in the following, whenever we refer to an 
angle we will take this as the center of the corresponding bin.
Note that there is only one case of a negative angle, the relation 
schematically drawn in the left-hand-side inset of Fig. 5a. 
The most frequently observed angle under
which the short-term MA intersects the horizontal is 67$^{\circ}$.
The angles are relatively uniformly distributed between 16$^{\circ}$ and 
85$^{\circ}$ with additional spikes at 43$^{\circ}$, 79$^{\circ}$
and 85$^{\circ}$.

The distribution of the angle $\beta$ under which the long-term MA 
intersects the horizontal at a $G$ cross is rather different from that
for the angle $\alpha$, as can be seen in Fig. 6b. 
This histogram is approximately symmetrical with respect to positive and 
negative values with maxima at -14$^{\circ}$ and at 7$^{\circ}$.
The distribution of the angle $\gamma$ between the two MAs at
a $G$ cross (see Fig. 6c) is rather
uniform in the interval [20$^{\circ}$,50$^{\circ}$] with a maximum at
26$^{\circ}$ and a number of low frequency occurrences at 29, 38 and 
46$^{\circ}$.

The distribution of angles under which the two MAs intersect the 
horizontal and with one another at $D$ cross is plotted in Fig. 6(d-f). 
The angle $\alpha$ of the short-term MA is mostly negative with 
one case of positive 60$^{\circ}$ which corresponds to the case 
sketched in the right-hand-side inset of Fig. 5b. 
At the time of crossings the $\alpha$ angles of the 
short-term MA are paired with $\beta$ angles whose distribution is 
plotted in Fig. 6e.
Therefore, the negative $\alpha$ angles and positive $\beta$ angles 
correspond to the schematic representation in Fig. 5b, while negative 
$\alpha$ and negative $\beta$ values represent the case sketched 
in the left-hand-side inset in Fig.  5b. 
The maximum value of the $\alpha$ angle is at -82$^{\circ}$. 
One is very unlikely to observe an angle of -47$^{\circ}$ in the data. 
The maximum $\beta$ value is equal to 7$^{\circ}$, so the long-term 
MA is most likely to be rather close
to the horizontal with a quasi-homogeneous spread over the interval
[-10$^{\circ}$, 10$^{\circ}$]. 
In comparison the spread of $\beta$ values at $G$ crosses 
reported above is a bit wider. 
The most likely angle $\gamma$ between the MAs at $D$
cross can have two values 23$^{\circ}$ and 59$^{\circ}$ 
with 33\% less chance to have a value of 47$^{\circ}$.

Clearly, the distribution of the slopes of the MAs at crosses as well
as the angles between them is very rich and far from random.  It remains
to be seen if similar patterns can be detected in other financial time
series and if their occurrence can be related to features of the spectrum.
However, it seems evident that more sophisticated prediction strategies
should take these variables into account.

Next, we have studied the distribution of the time intervals between 
successive [$G$,$G$], [$D$,$D$], [$G$,$D$], and [$D$,$G$] crosses 
for the S\&P500 closing price signal in two distinct time intervals chosen
because they differ strongly in their investment environment.  
These are the time periods from Jan 1, 1980 to Dec 31, 1990 (see
Fig. 7(a-d)) and from Jan 1, 1991 to Dec 31, 2001 (see Fig. 7(e-h)),
the former generally characterized by a rather ``sluggish" economy,
the latter mainly reflecting the ``bull market" of the 1990s.

While the shortest interval between successive [$G$,$G$] crosses in the
first case appears to be 3 days, during the second period it is twice 
as long, i.e. 6 days. 
The maximum of the distribution is observed to occur at 9 and 13 for the first
period and at 17 days for the second period, and is thus not drastically
different
between the two periods. The most probable time interval between successive 
[$D$,$D$] crosses during the first period is 19 days. Time intervals of
similar
length: 20-22 days, e.g. approximately one market month, are very unlikely to 
occur during the second period, whose maxima are at 16 and 28 days.   Thus, 
downturns tended to occur more frequently in the 1980s than in the 1990s.  
A measure of the time necessary for the market to recover is the interval 
between [$G$,$D$] crosses. Its maximum is at 4 days during the first period. 
In sharp contrast, the second period is characterized by a high frequency of 
occurrence of [$G$,$D$] crosses for time intervals between 2 to 6 days, 
indicating that the market tended to recover faster during the second 
observation interval.

The way the market is going down from a $D$ to $G$ cross is reflected in the
distribution of time intervals between consecutive  $D$ and $G$ crosses. 
During both periods, the most likely interval between successive $D$ and 
$G$ crosses is 4 days. 

\section{Minima-Maxima distribution}

As can be noticed in Fig. 2, the occurrences of $G$ and $D$
crosses are preceded by turnovers of the trend of the MAs. 
Thus the distribution of time intervals between successive 
minima and maxima can be calculated in order to provide 
information on the dynamics of the market. 
To this end for the S\&P500 signal studied here the PDFs 
of the time interval between successive maxima [$M_1,M_2$], 
minima [$m_1,m_2$], maxima-minima [$M_1,m_1$], and minima-maxima
[$m_1,M_1$], are plotted in Fig. 8(a-h) for the short-term, 
$y_{week}$, and long-term, $y_{month}$, MAs. 
The x-axis of all data in Fig. 7(a-h) is chosen in such a way 
that the bin of each x-axis is equal to one day. Also the 
ticks are associated with the middle of the bin, i.e.  to 0.5. 
For the purpose of comparison the insets of Fig. 8(c-h) 
have their x- and y-axis limits the same as those of Fig. 8(a,b).  
It is of interest to note that the PDFs of [$M_1,M_2$] and 
[$m_1,m_2$] for the short-term MA, $y_{week}$, are very similar, 
which is not surprising, both with a maximum at 2 days. 
In contrast, the [$m_1,m_2$] PDF for the long-term MA, $y_{month}$, 
exhibits a maximum at 2 days and a rather homogeneous distribution 
for time intervals between 3 and 11 days, while the respective 
maxima-minima [$M_1,m_1$] PDF has the appearance of an exponential one,
although the scarcity of data does not allow us to make this statement
rigorous.

We also checked the frequency of appearance of combinations of 
positive and negative increments of the MAs. 
Let a positive increment be marked by 1 and a negative 
increment be marked by -1. 
Then we test the frequency of occurrence of time intervals between 
two consecutive triplets (1 1 -1) and between two consecutive 
triplets (-1-1 1) for the two MAs of interest. 
These data are shown in Fig. 9 for the weekly and monthly MA
and the entire 22-year period. The increment of the histogram is 1~day.
The weekly MA $y_{week}$ has a maximum of [(1 1 -1),(1 1 -1)] 
time interval at 9 days.  
The maximum of the [(-1-1 1),(-1-1 1)] interval for 
$y_{week}$ is at 8 days with very low probability of
occurrence at 4 days time interval. 
The maxima for [(1 1 -1),(1 1 -1)] is at 5~days and for [(-1-1 1),(-1-1 1)] 
at 3~days for the monthly MA $y_{month}$, 
while it is very unlikely that (1 1 -1) will be followed by 
(1 1 -1) after 21~days. There are also many time intervals that have 
zero probability to occur for [(-1-1 1),(-1-1 1)]. They
are 20, 21, 30, 36, 39, 48, 50, 51, 53, 56, 61 days, to mention
some of them. 

\section {Conclusion}

We presented a set of quantitative measures of two MAs of
the S\&P500 daily closing price signal and of their relative position to
one another. Since different MAs represent the trend of the signal
looked upon from different investment horizons, the measures of 
their relative geometry provide information on the change of the trend of 
the signal. The distribution of the time interval between 
$DD,GG,DG$ and $GD$ crosses, together with that of the maxima
and minima in the MA, and the distribution of the
angle between the two MAs at the crosses can serve as ingredients 
for an investment strategy that enriches the classical technical 
analysis with parameters of the dynamics of the signal trend, i.e.
of the momentum of S\&P500 closing price. 
We emphasize, that these conclusions should not be taken as an endorsement
of chartist strategies or technical analysis in general.  Whether it is 
possible to base a successful trading strategy on the rather subtle patterns 
observed here remains to be seen and has not been addressed in this paper.

\vskip 0.5cm {\bf Acknowledgements} We thank Dr. M. Ausloos for 
critical comments and suggestions on this work.

\newpage \begin{figure}[ht] \begin{center} 
\leavevmode \epsfysize=6cm
\epsffile{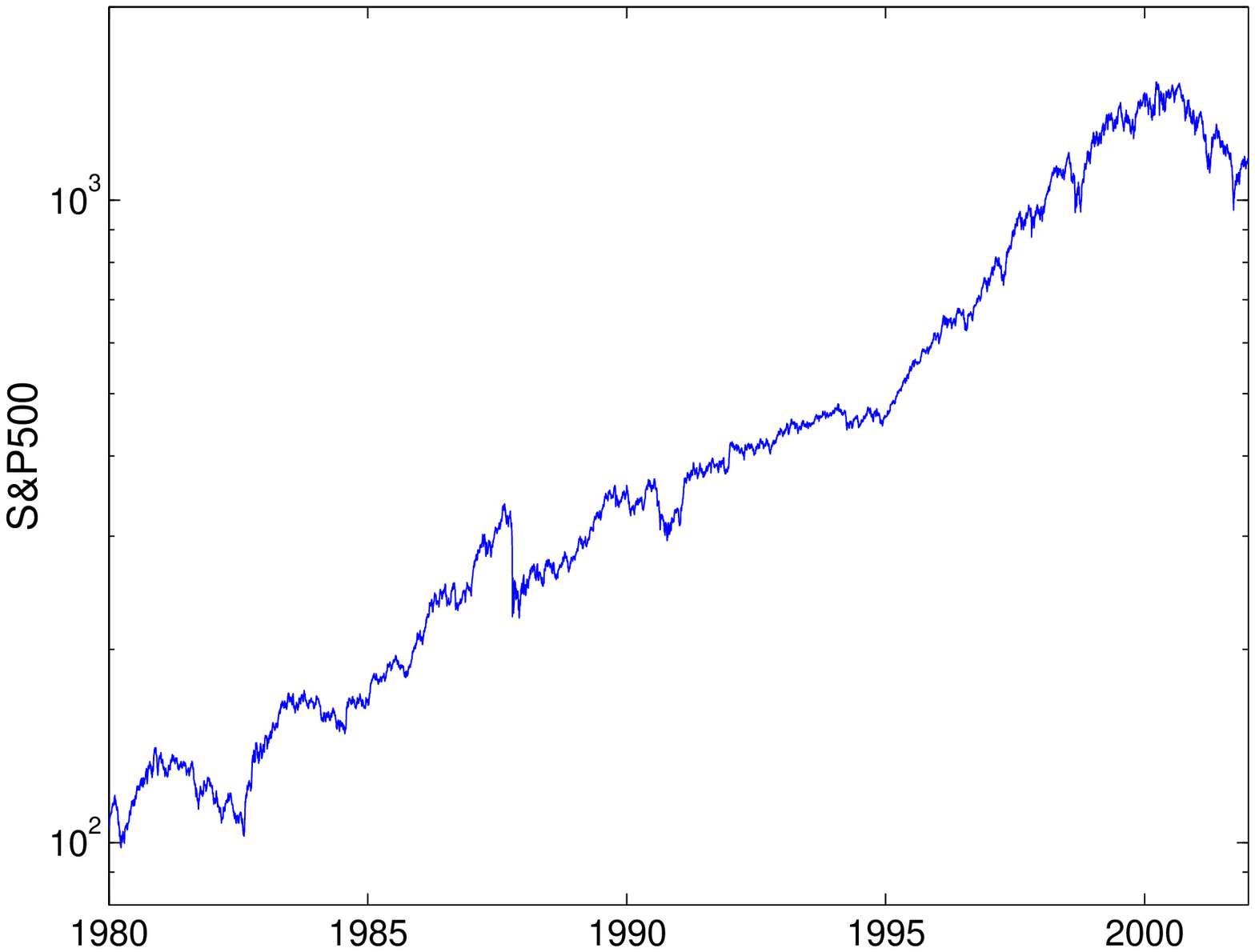} 
\vfill \leavevmode \epsfysize=6cm \epsffile{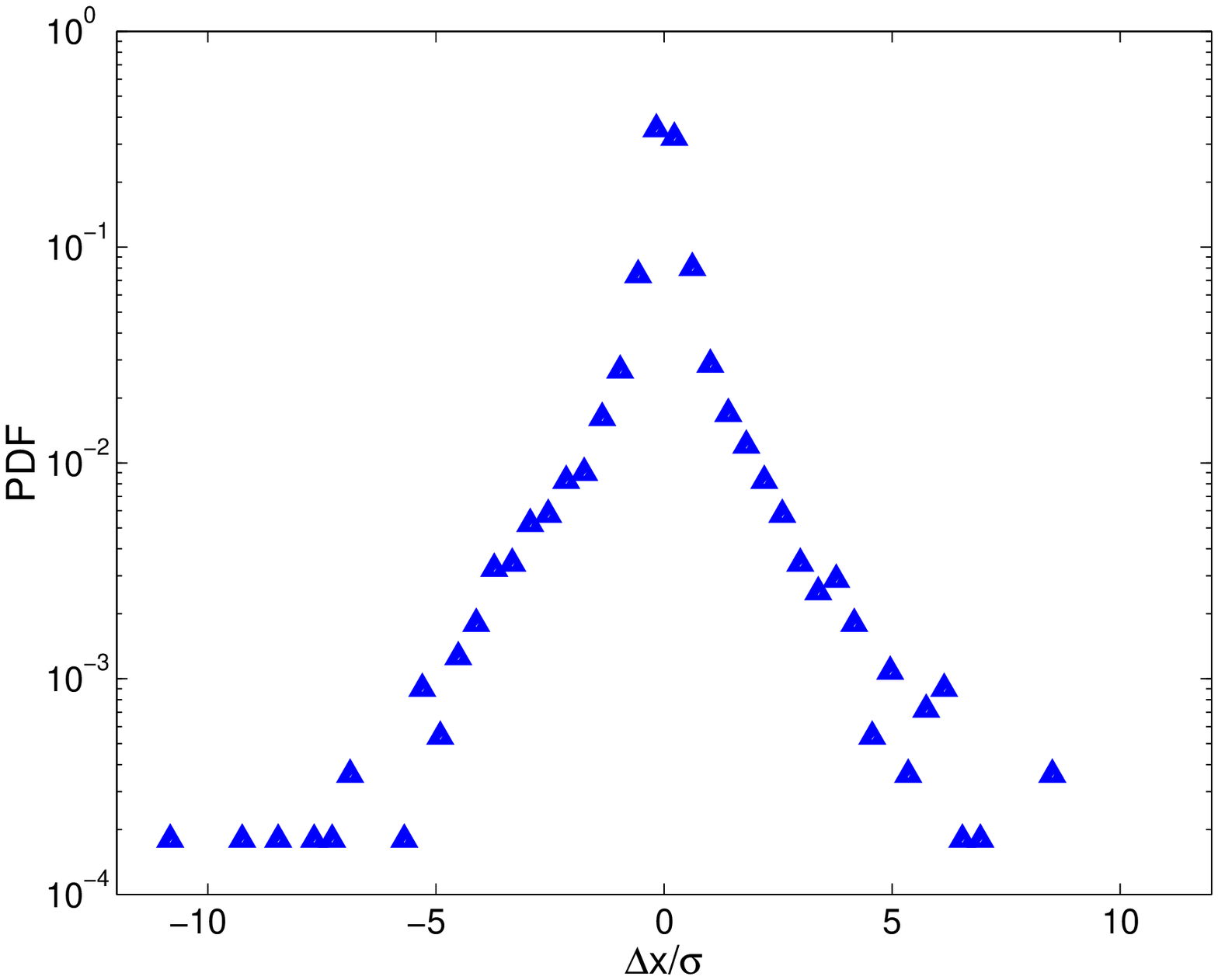}
\end{center} \caption{(a) S\&P500 daily closing value signal 
between Jan. 01,
1980 and Dec. 31, 2001, i.e. 5556 data points. (b) Probability density 
function of fluctuations of S\&P500 signal.\cite{yahoosp500} } \end{figure}

\newpage \begin{figure}[ht] \begin{center} 
\leavevmode \epsfysize=6cm
\epsffile{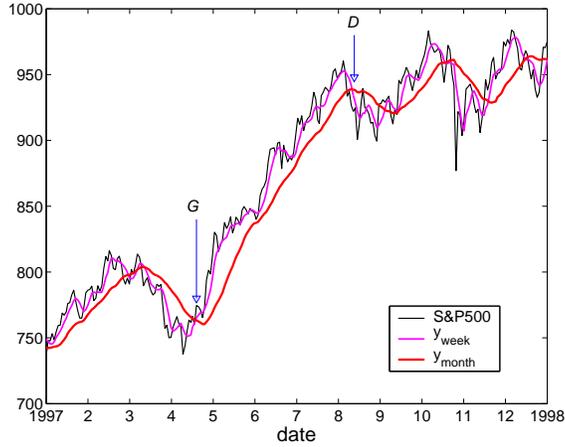} 
\end{center} \caption{Typical 
S\&P500 daily
closing value signal between Jan. 01, 1997 and Dec. 31, 1998, with two moving
averages, $y_{week}$ and $y_{month}$ for $T_1=5$~days and 
$T_2=21$~days. $D$ and $G$ crosses, are defined in the text.} \end{figure}

\begin{figure}[ht] \begin{center} 
\leavevmode \epsfysize=6cm
\epsffile{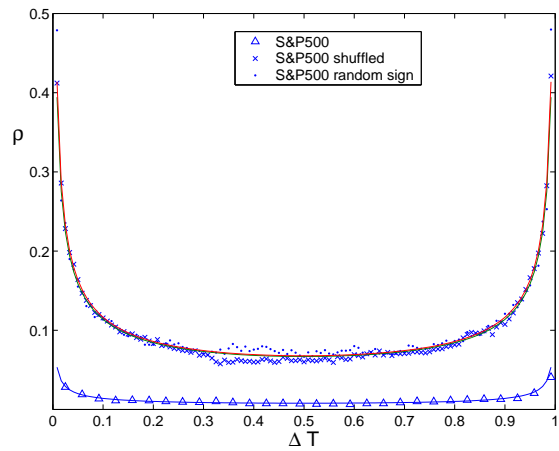} 
\end{center} \caption{The density $\rho$ of crossing points as a
function of the relative difference $\Delta T$ with $T_2=120$~days. 
S\&P500 daily
closing value signal (triangles), S\&P500 shuffled values signal (x), 
and S\&P500 with
randomized sign and preserving the amplitude of the signal (dots). 
Continuous lines
are fits to Eq. (2). } \end{figure}

\newpage \begin{figure}[ht] \begin{center} 
\leavevmode \epsfysize=6cm
\epsffile{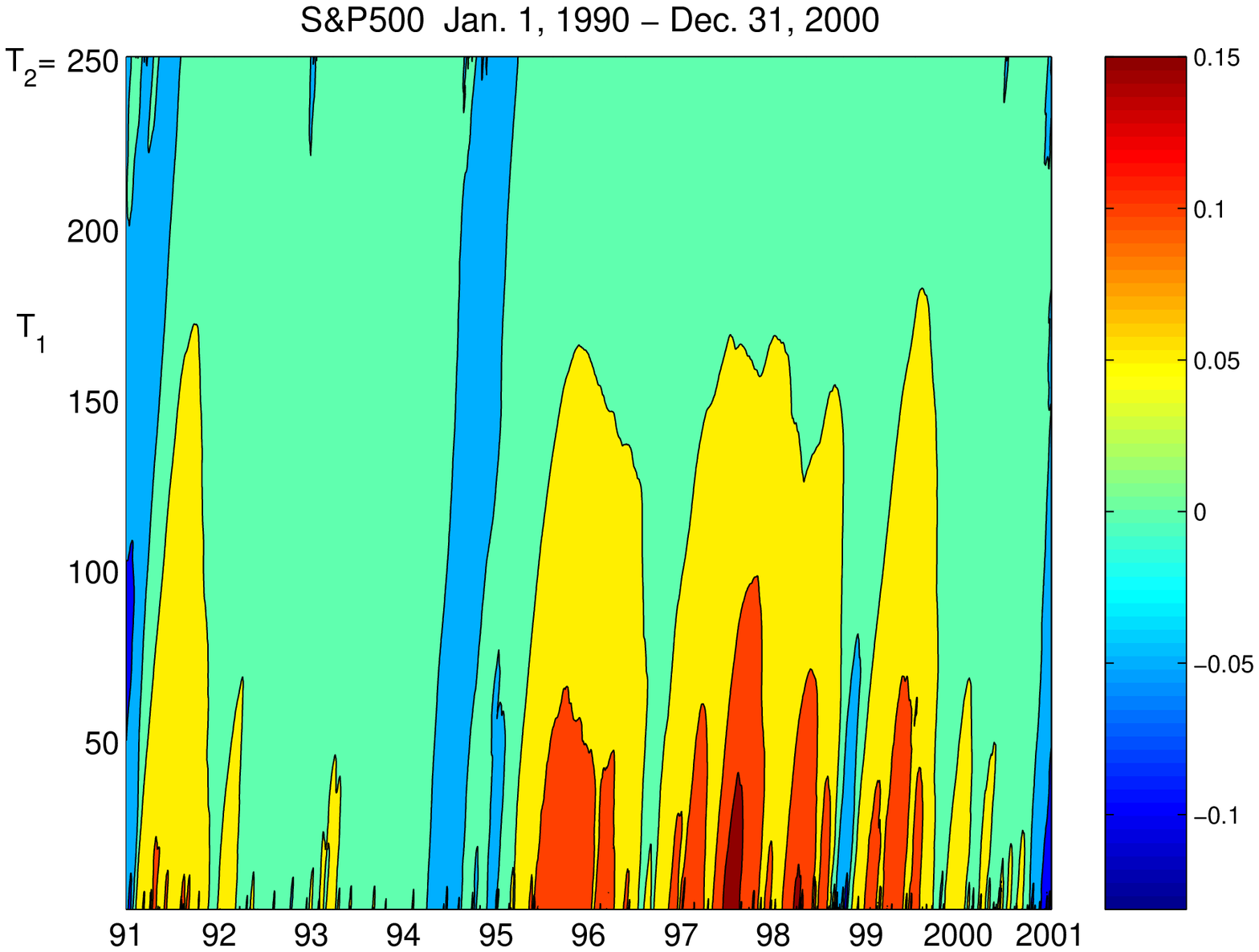} 
\vfill \leavevmode \epsfysize=6cm \epsffile{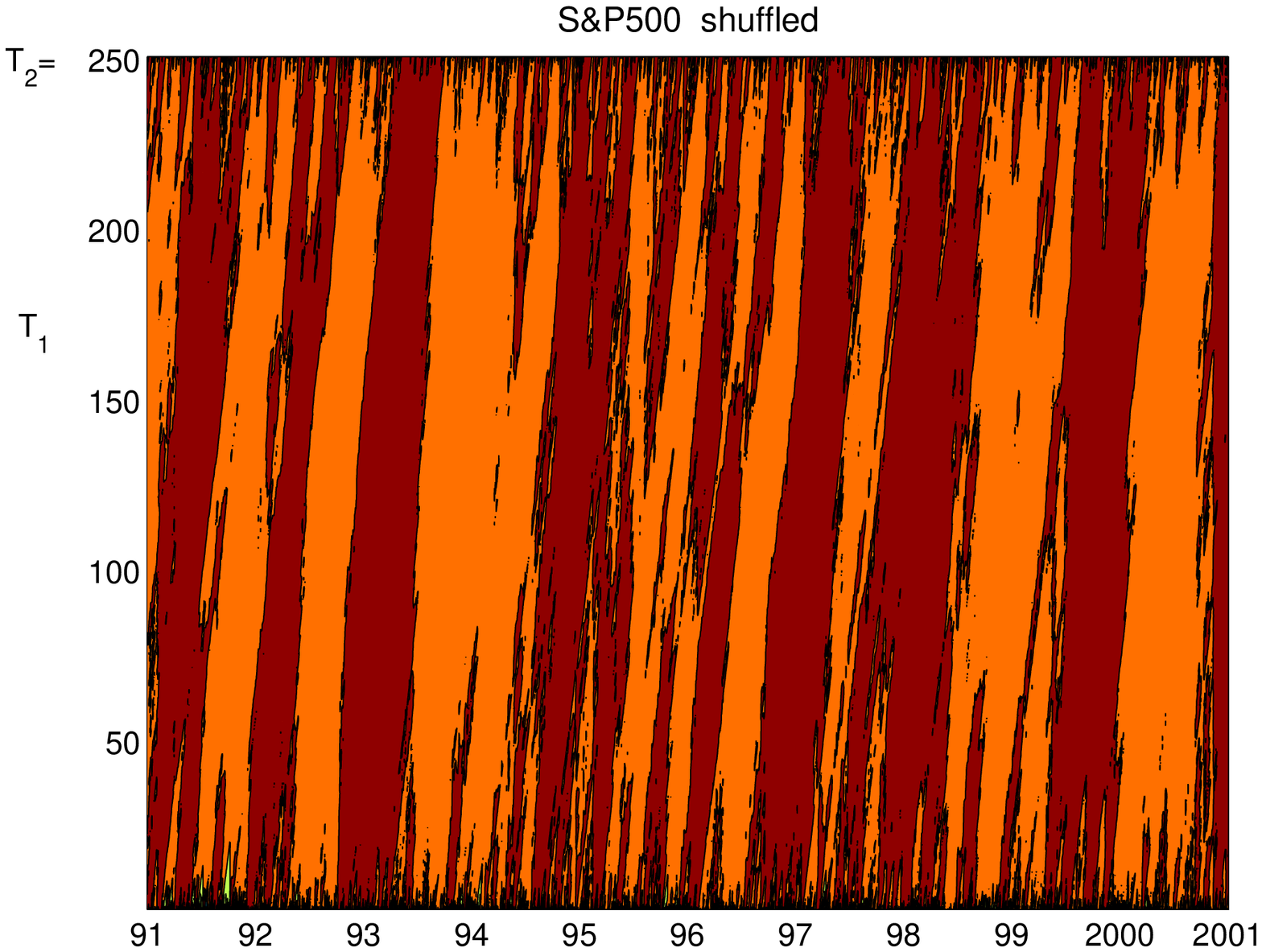}
\end{center} \caption{(a) The spectrum of moving averages for the
S\&P500 daily closing value signal data between Jan. 01, 1990 and 
Dec. 31, 2000.
The y-axis corresponds to $T_1$. The long term is fixed to $T_2=$252, i.e. one
market year. (b) The spectrum of moving averages for the S\&P500 shuffled 
signal. 
The grey levels correspond to the relative distance $\delta$ between the 
two moving averages. The darker the grey level the larger the difference.} 
\end{figure}

\newpage \begin{figure}[ht] \begin{center} 
\leavevmode \epsfysize=6cm
\epsffile{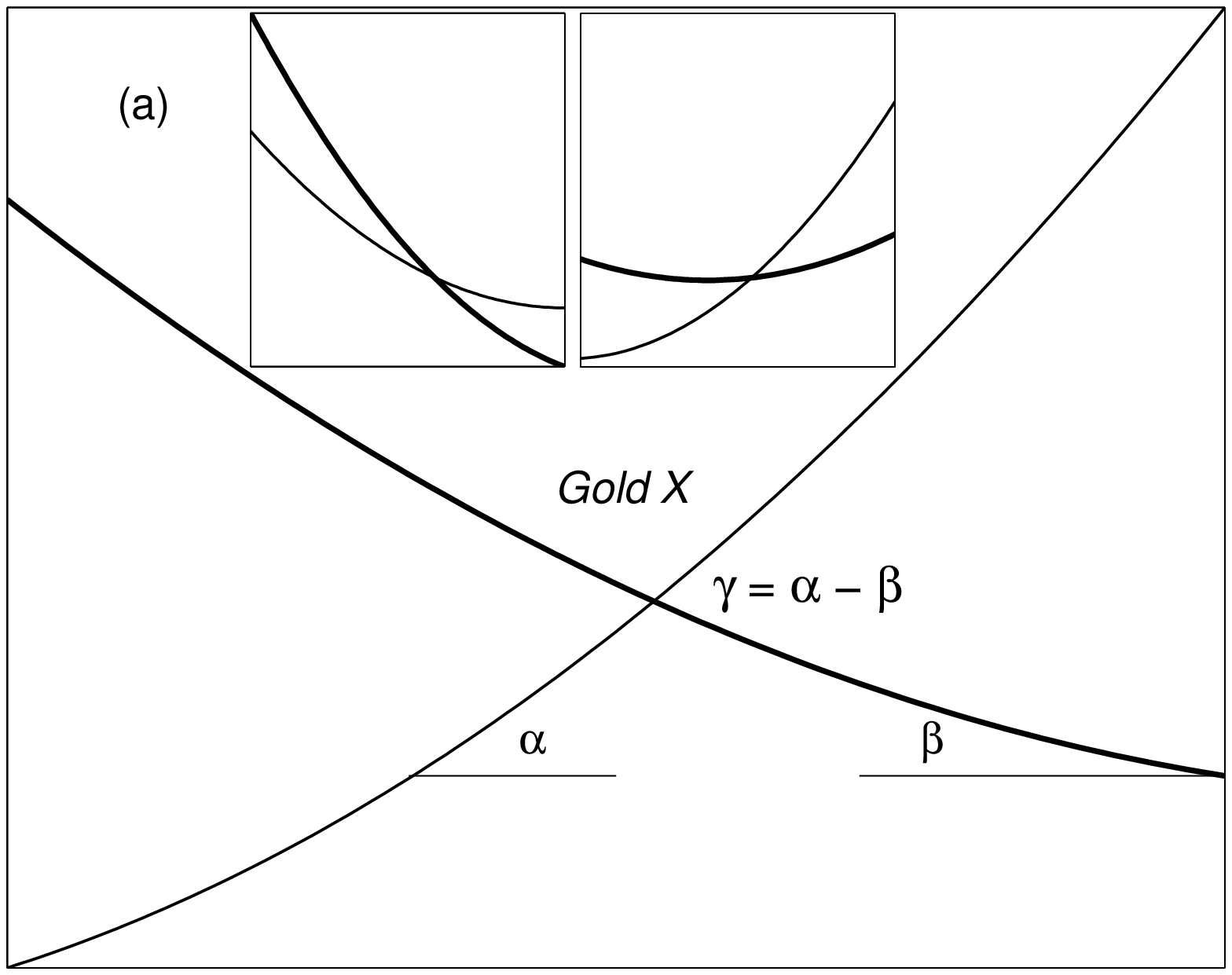} 
\vfill \leavevmode \epsfysize=6cm \epsffile{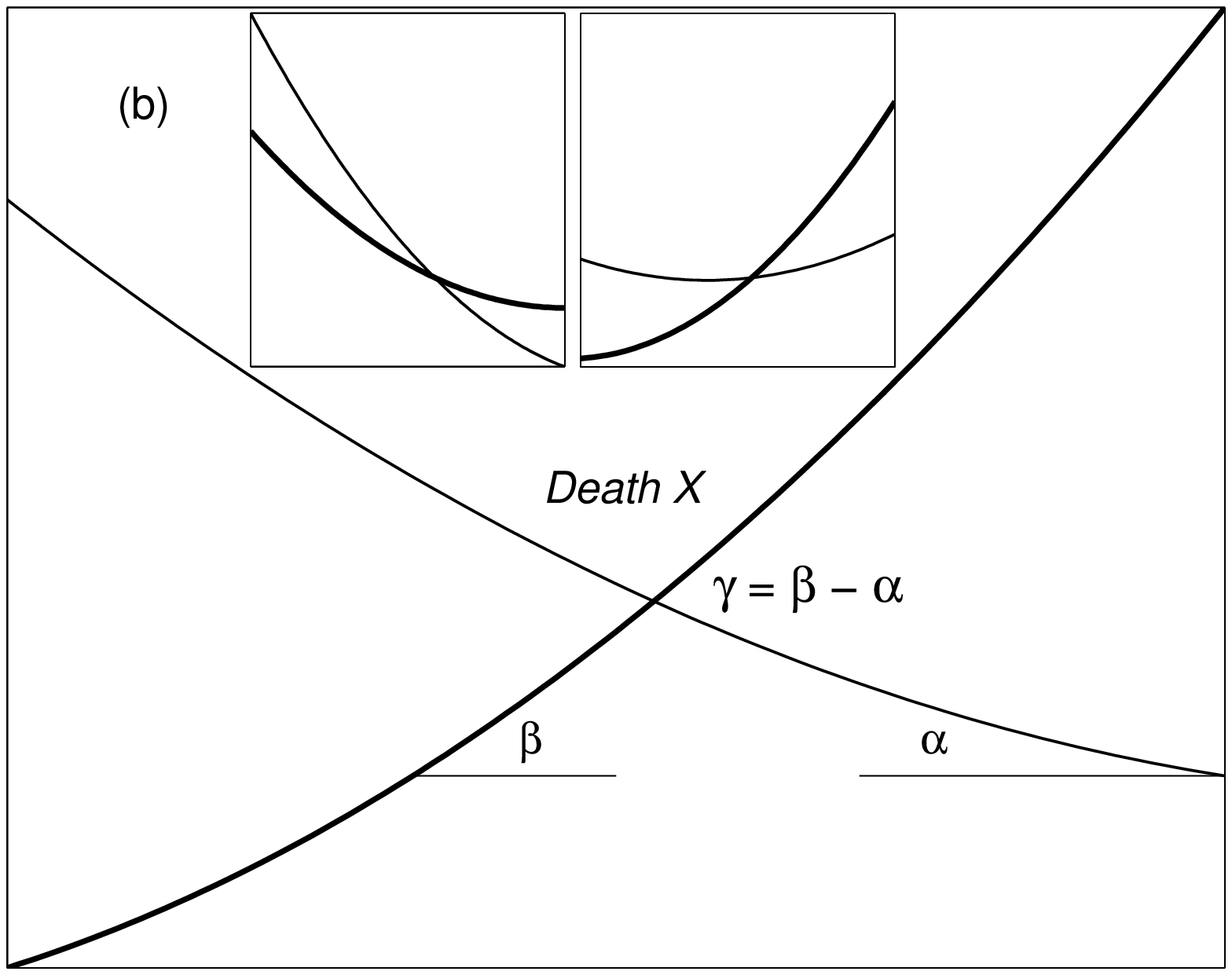}
\end{center} \caption{(a) Schematic presentation of a {\it Gold cross}
between two moving averages. (b) Schematic presentation of a {\it Death cross}
between two moving averages. The thicker lines correspond to the longer-term
moving average. Noted are the angles between the horizontal and the short-
($\alpha$) and long-term ($\beta$) moving averages; $\gamma$ denotes the angle
between both moving averages.} \end{figure}

\newpage \begin{figure}[ht] \begin{center} 
\leavevmode \epsfysize=4cm
\epsffile{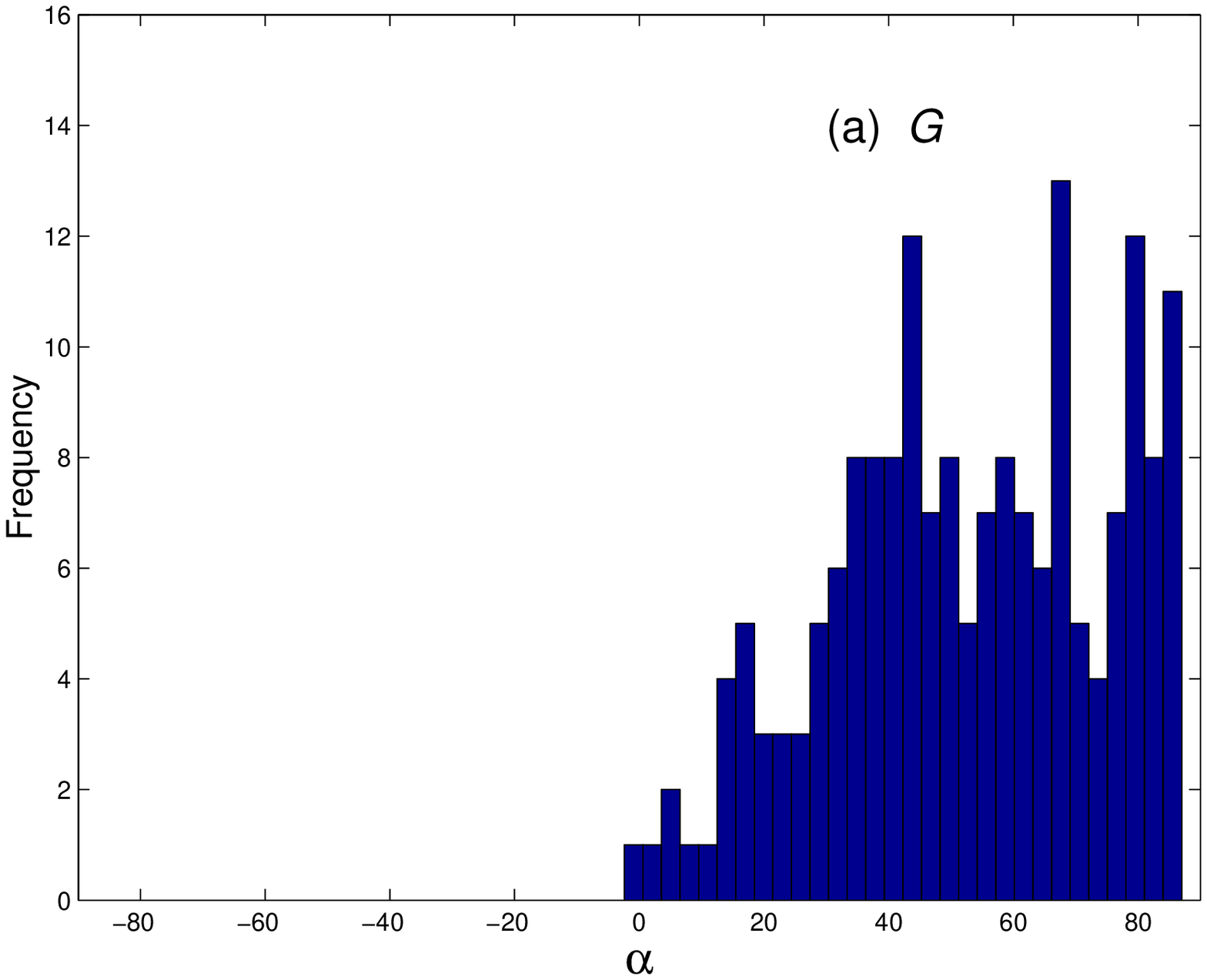} 
\hfill \leavevmode \epsfysize=4cm \epsffile{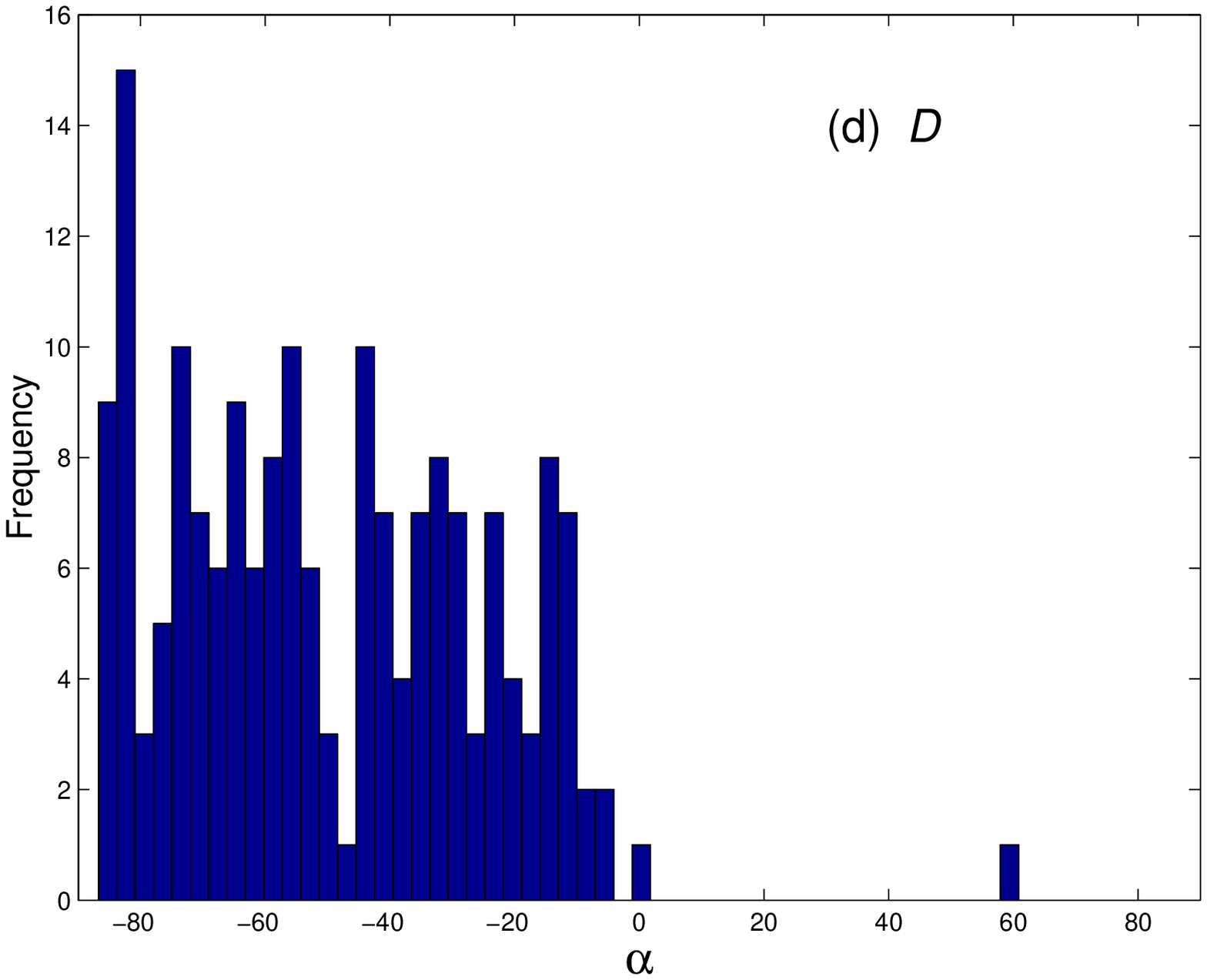}
\vfill \leavevmode \epsfysize=4cm \epsffile{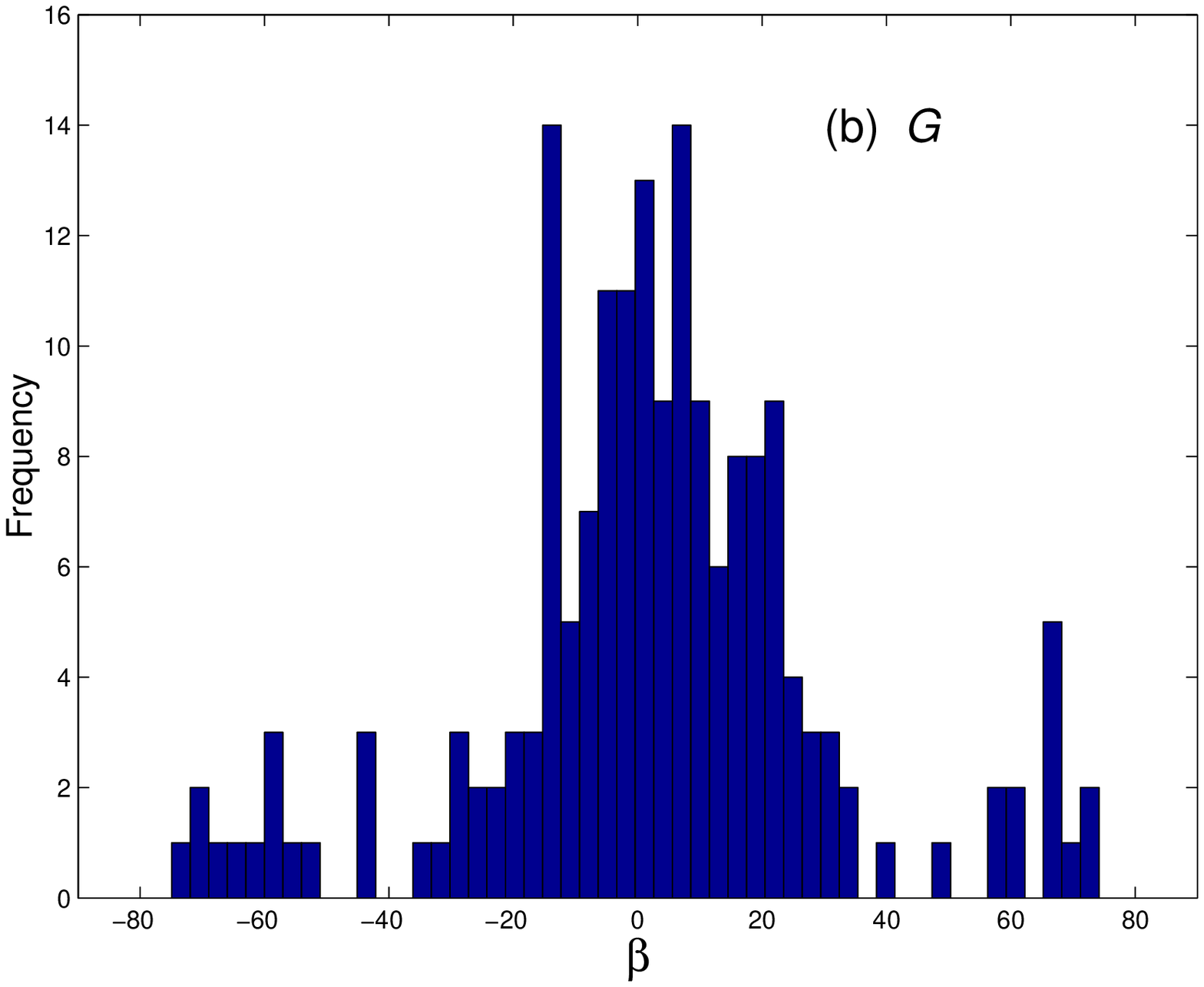}
\hfill \leavevmode \epsfysize=4cm \epsffile{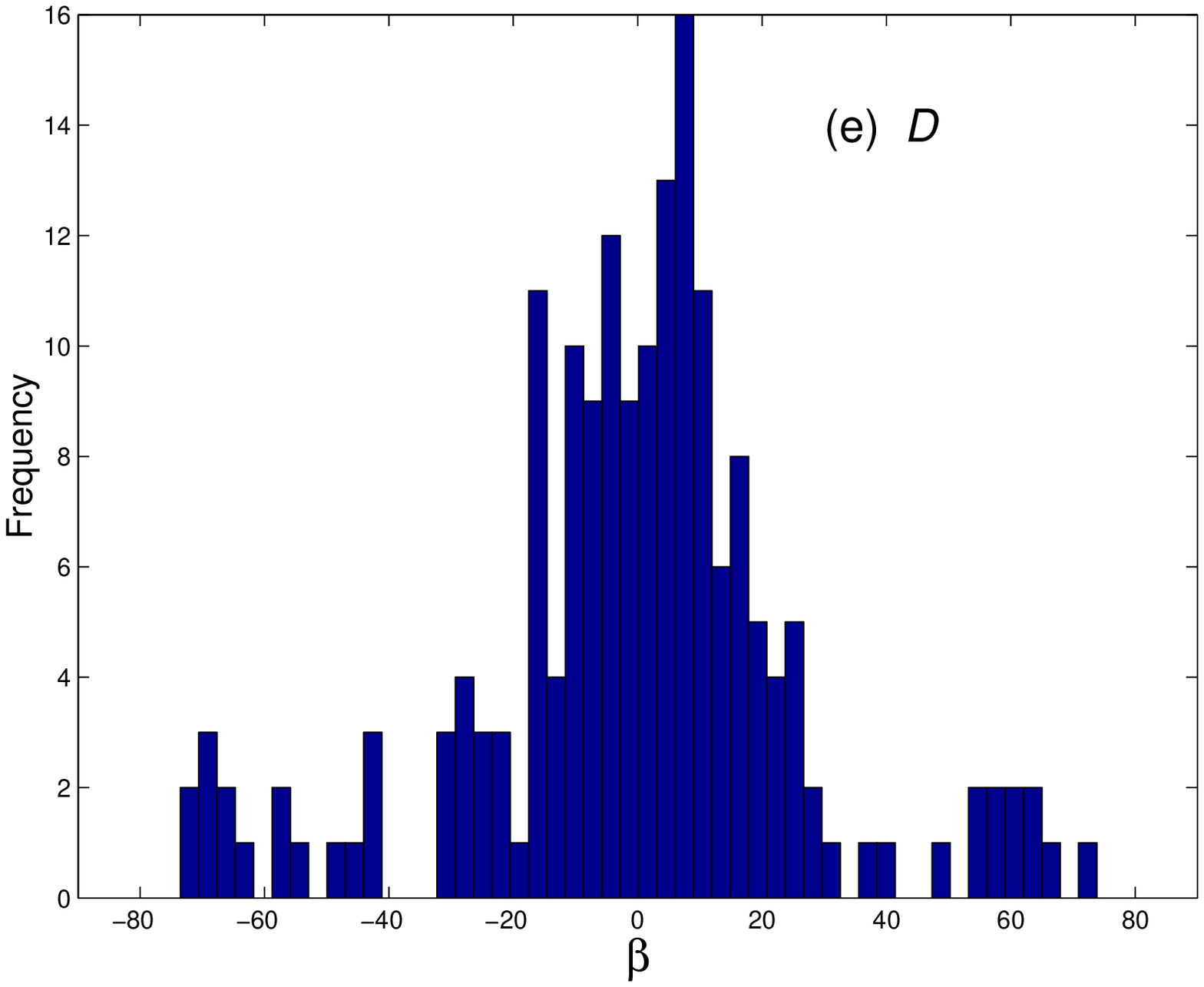}
\vfill \leavevmode \epsfysize=4cm \epsffile{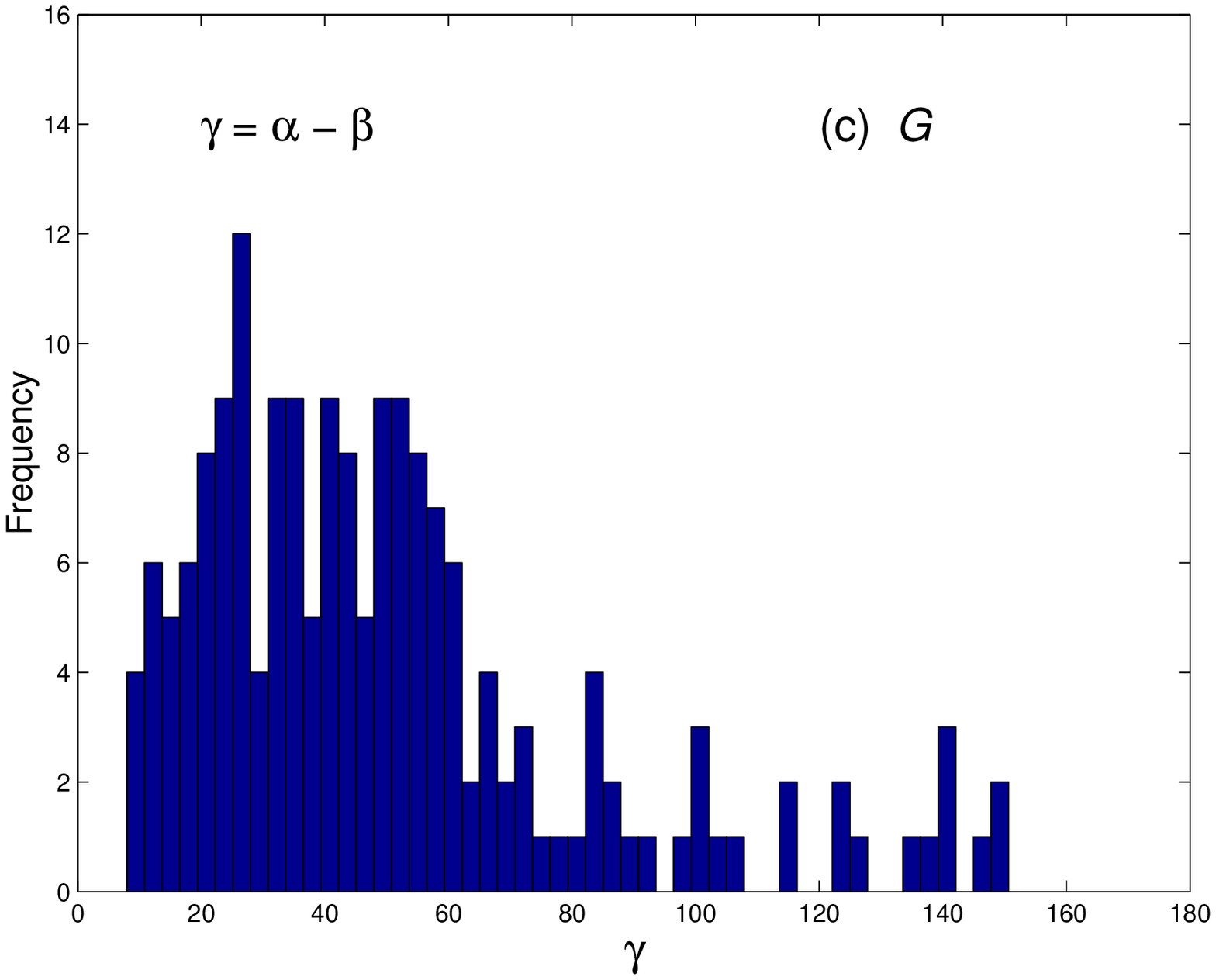}
\hfill \leavevmode \epsfysize=4cm \epsffile{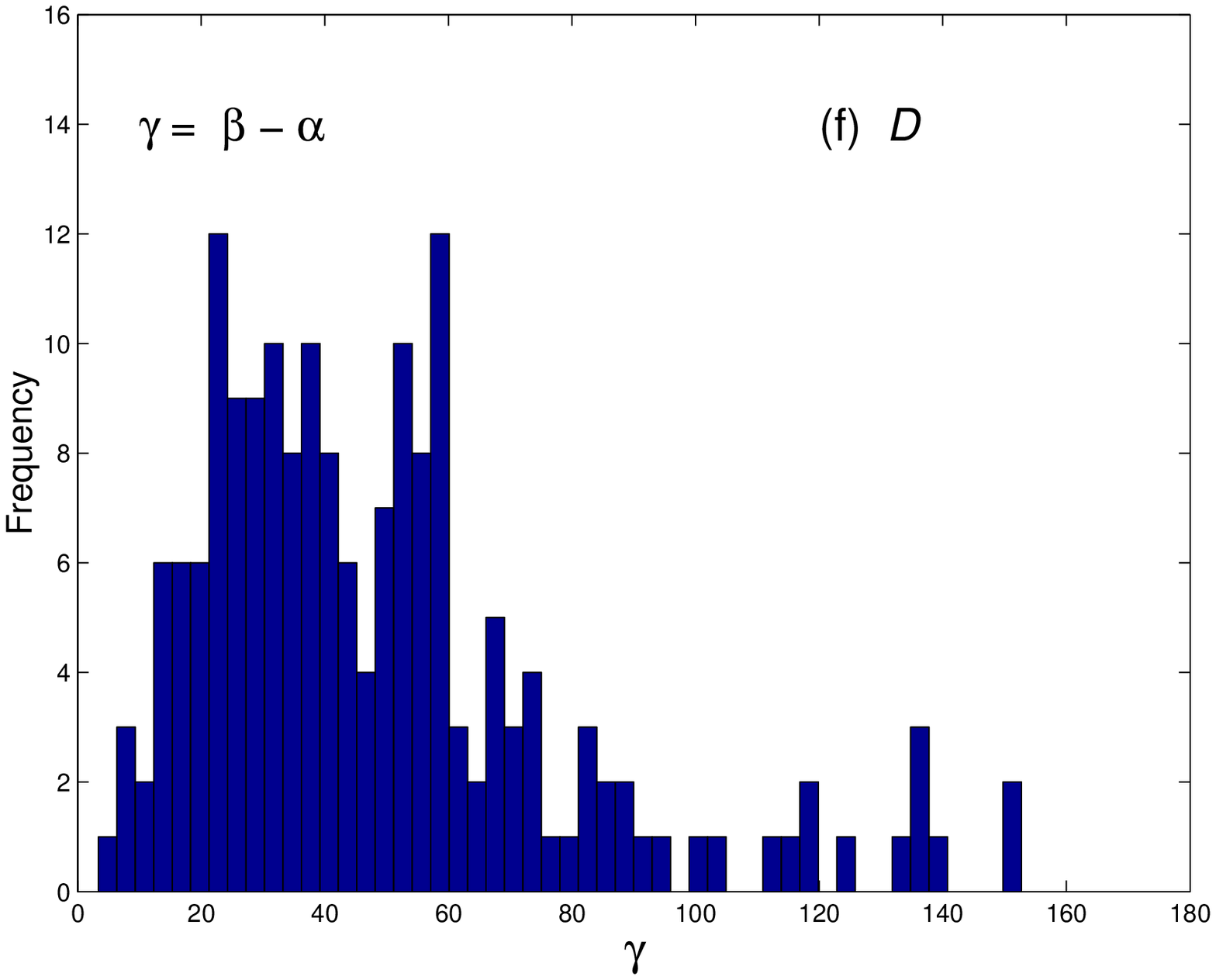}
\end{center} \caption{Frequency of the angle between the 
horizontal and
the short-term moving average $\alpha$: (a) for $G$ crosses and (d) for
$D$ crosses; between the horizontal and the long-term moving 
average $\beta$:
(b) for $G$ crosses and (e) for $D$ crosses; between the two moving
averages $\gamma$: (c) for $G$ crosses and (f) for $D$ crosses. Data
considered are the S\&P500 closing price from Jan 1, 1980 to Dec 31, 2001.
} \end{figure}

\newpage \begin{figure}[ht] \begin{center} \leavevmode \epsfysize=3.5cm 
\epsffile{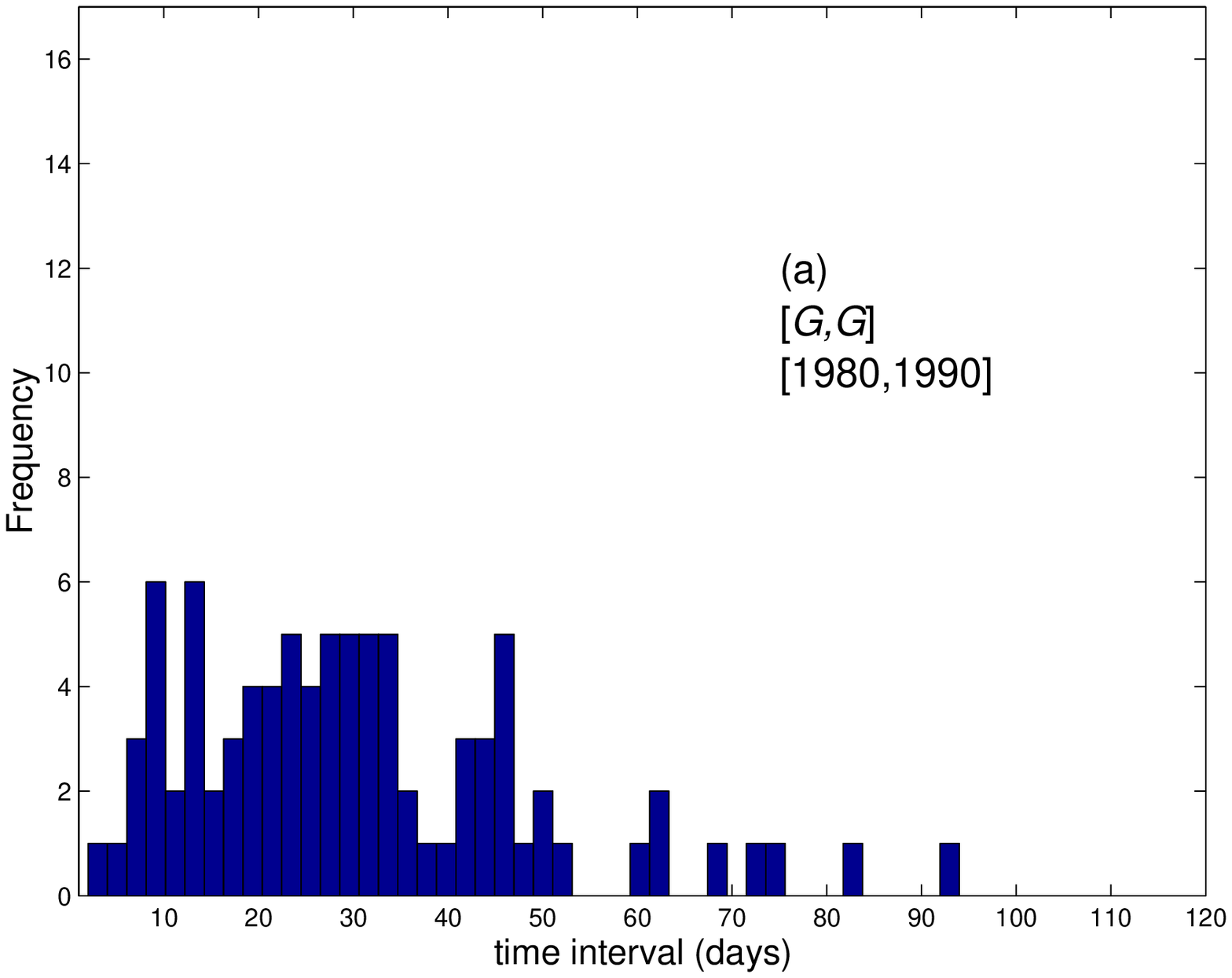}
\hfill \leavevmode \epsfysize=3.5cm \epsffile{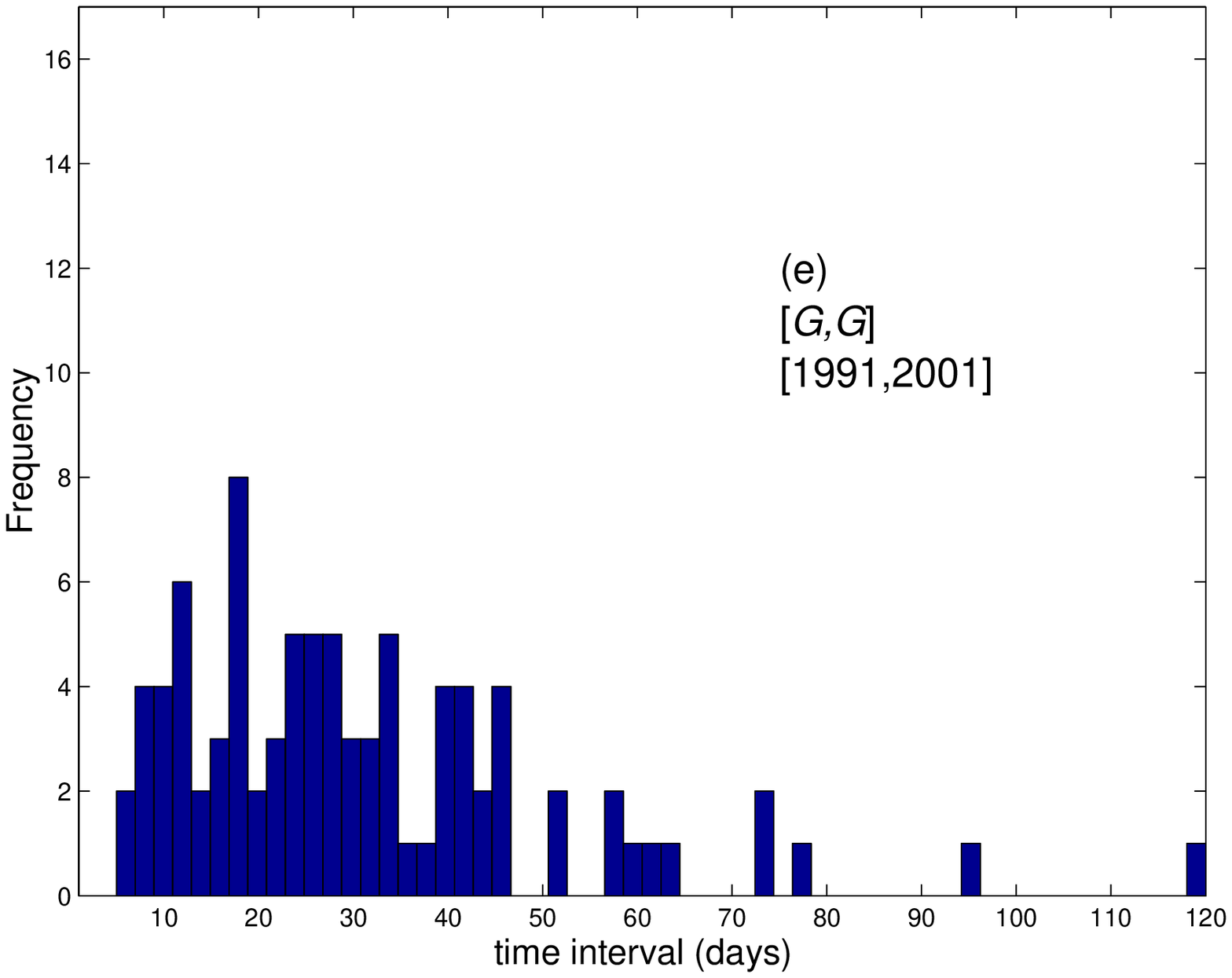}
\vfill \leavevmode \epsfysize=3.5cm \epsffile{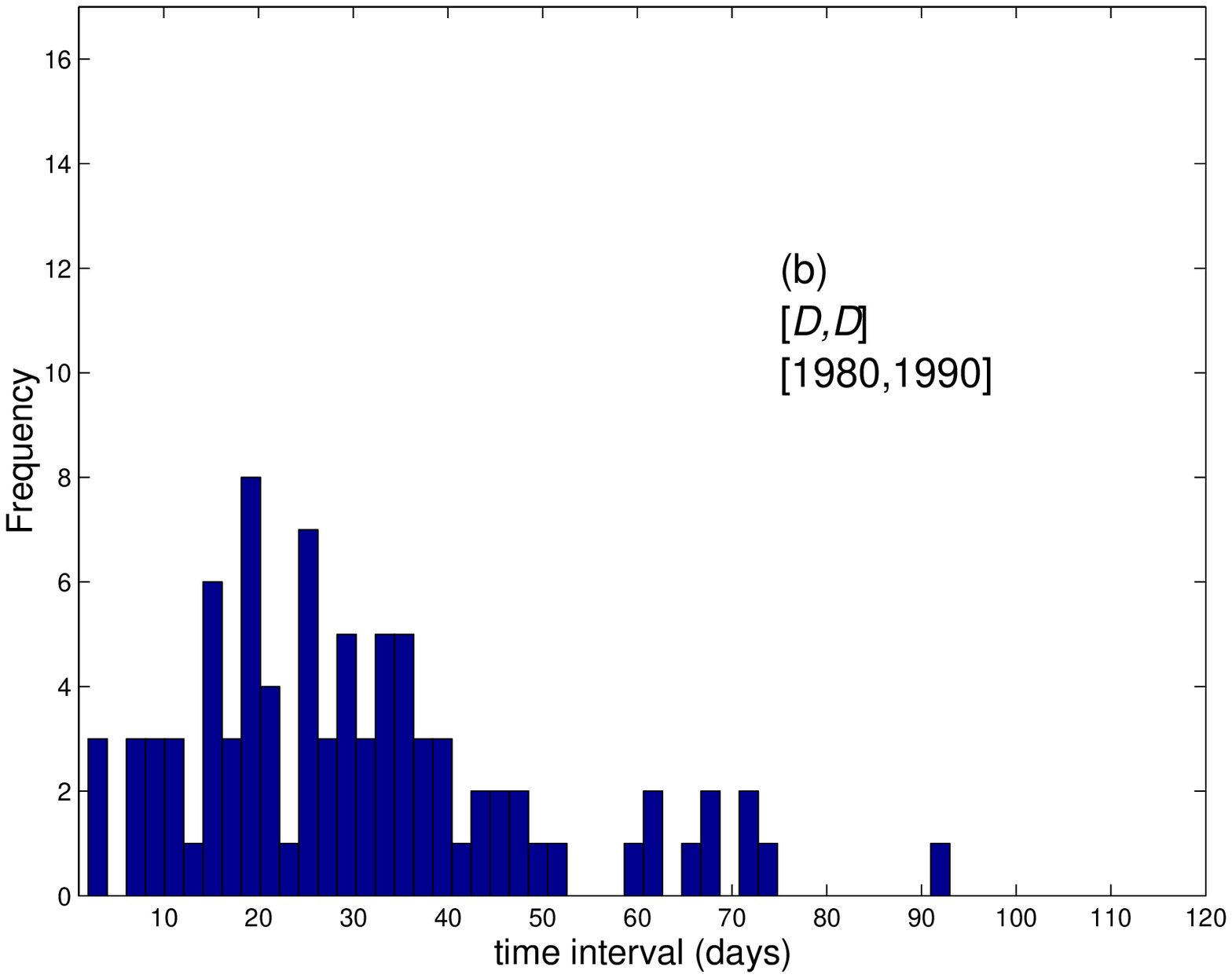}
\hfill \leavevmode \epsfysize=3.5cm \epsffile{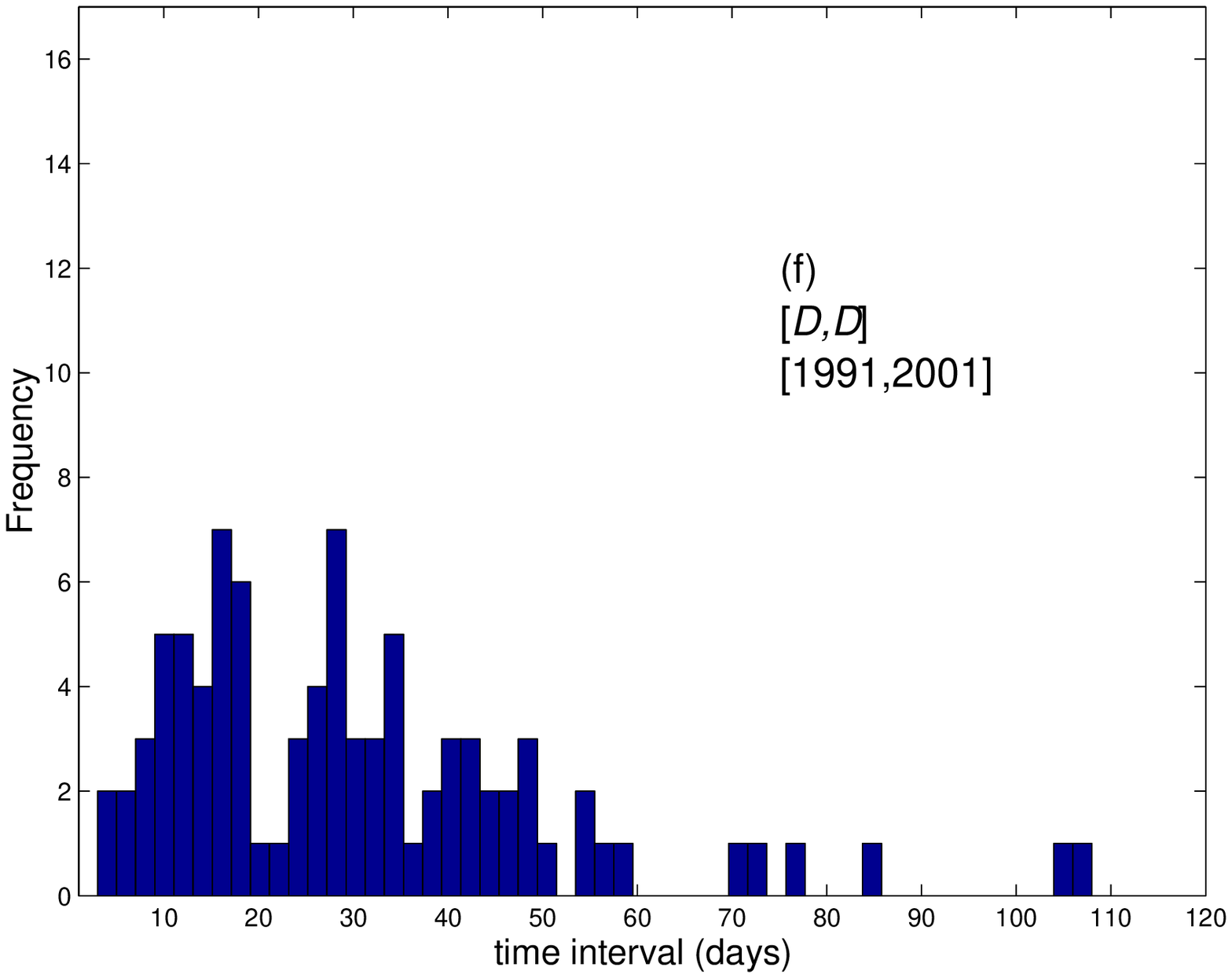}
\vfill \leavevmode \epsfysize=3.5cm \epsffile{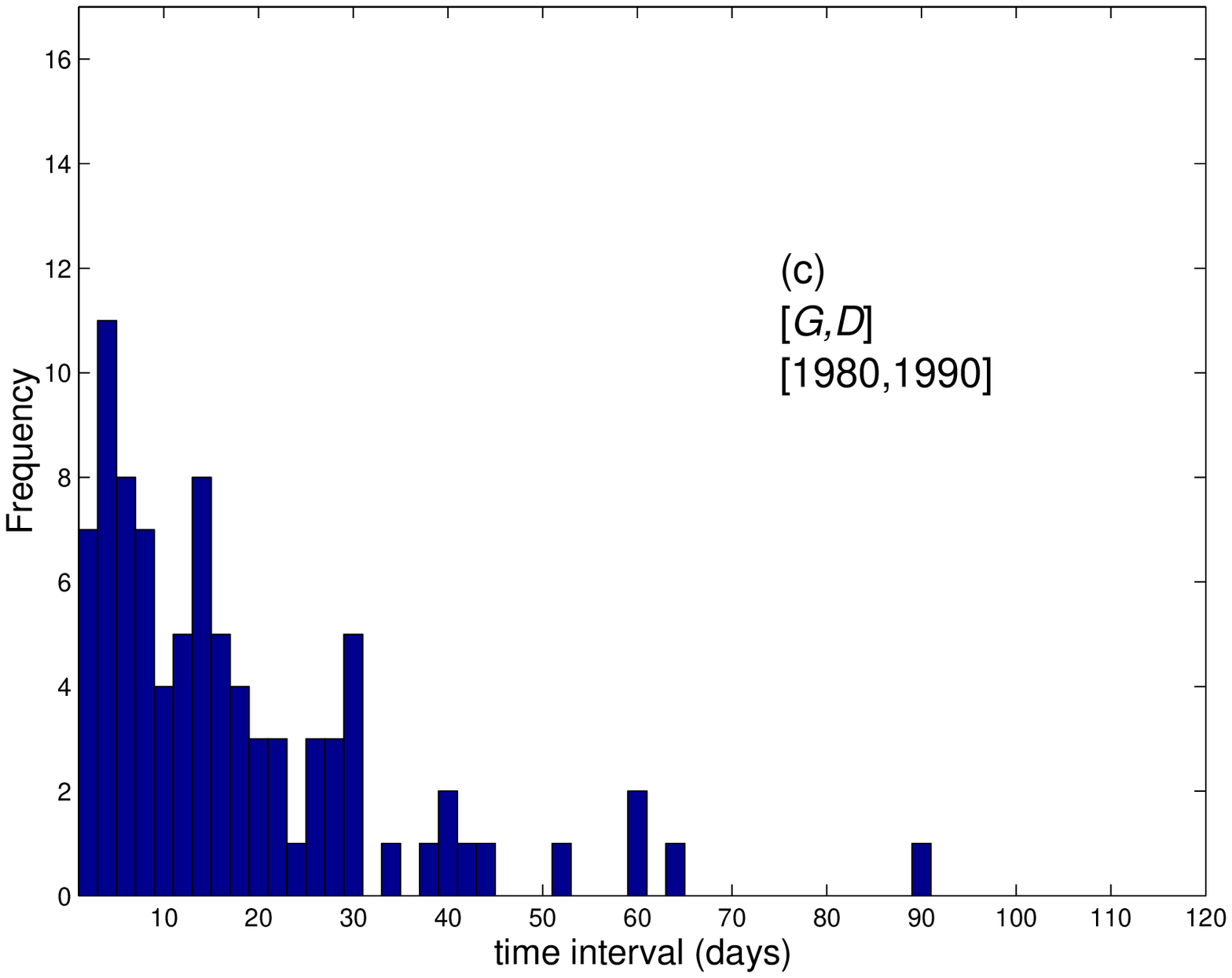}
\hfill \leavevmode \epsfysize=3.5cm \epsffile{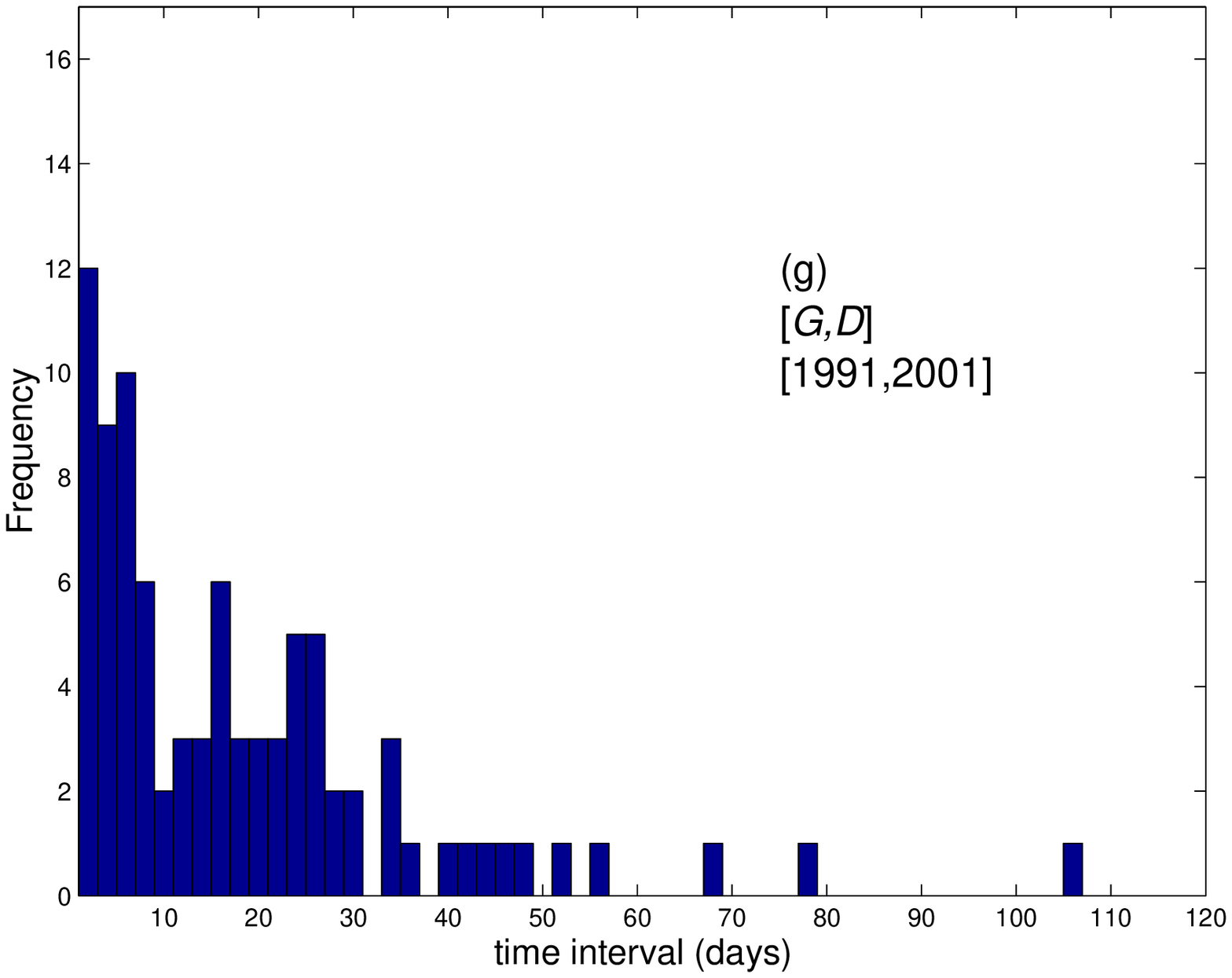}
\vfill \leavevmode \epsfysize=3.5cm \epsffile{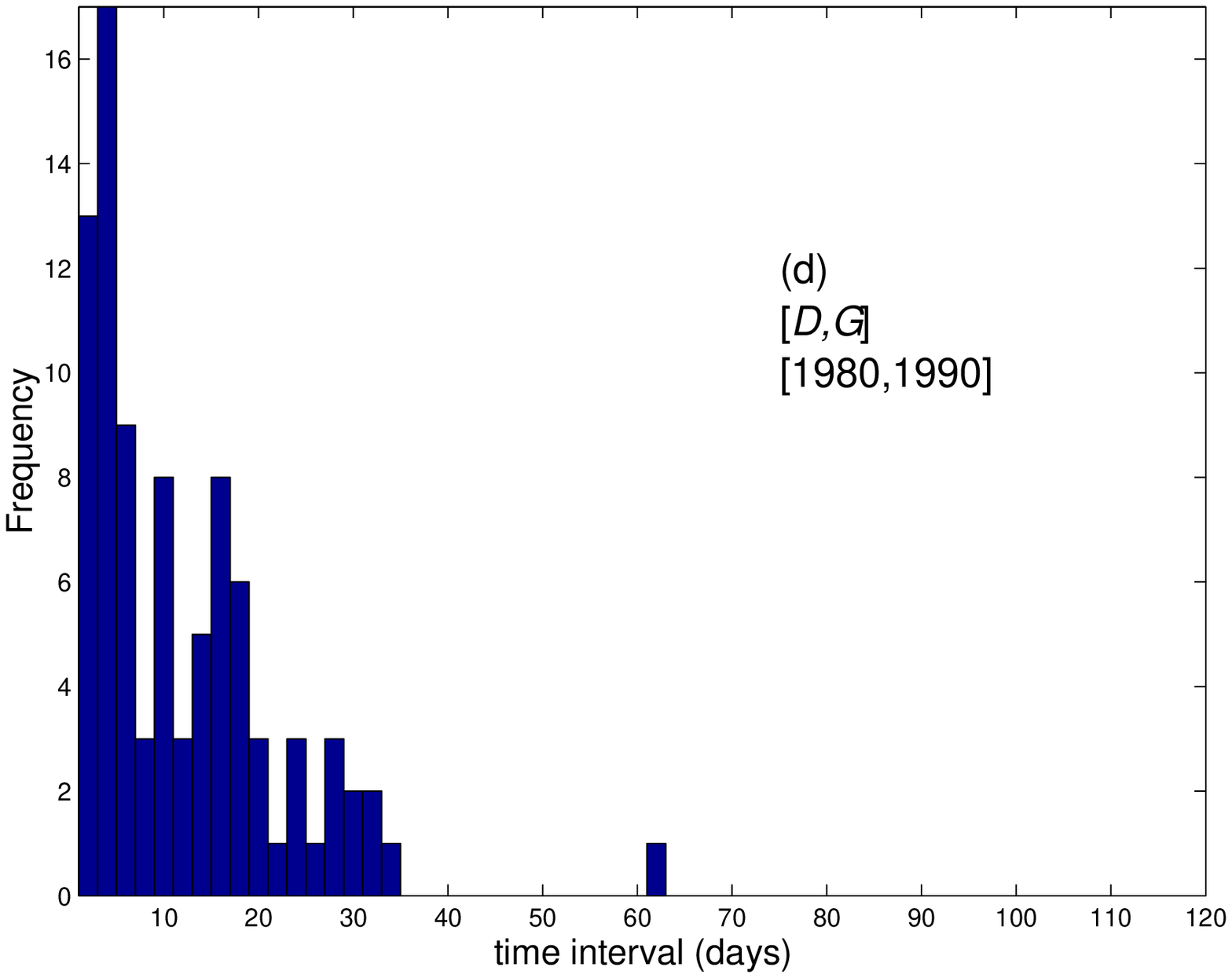}
\hfill \leavevmode \epsfysize=3.5cm \epsffile{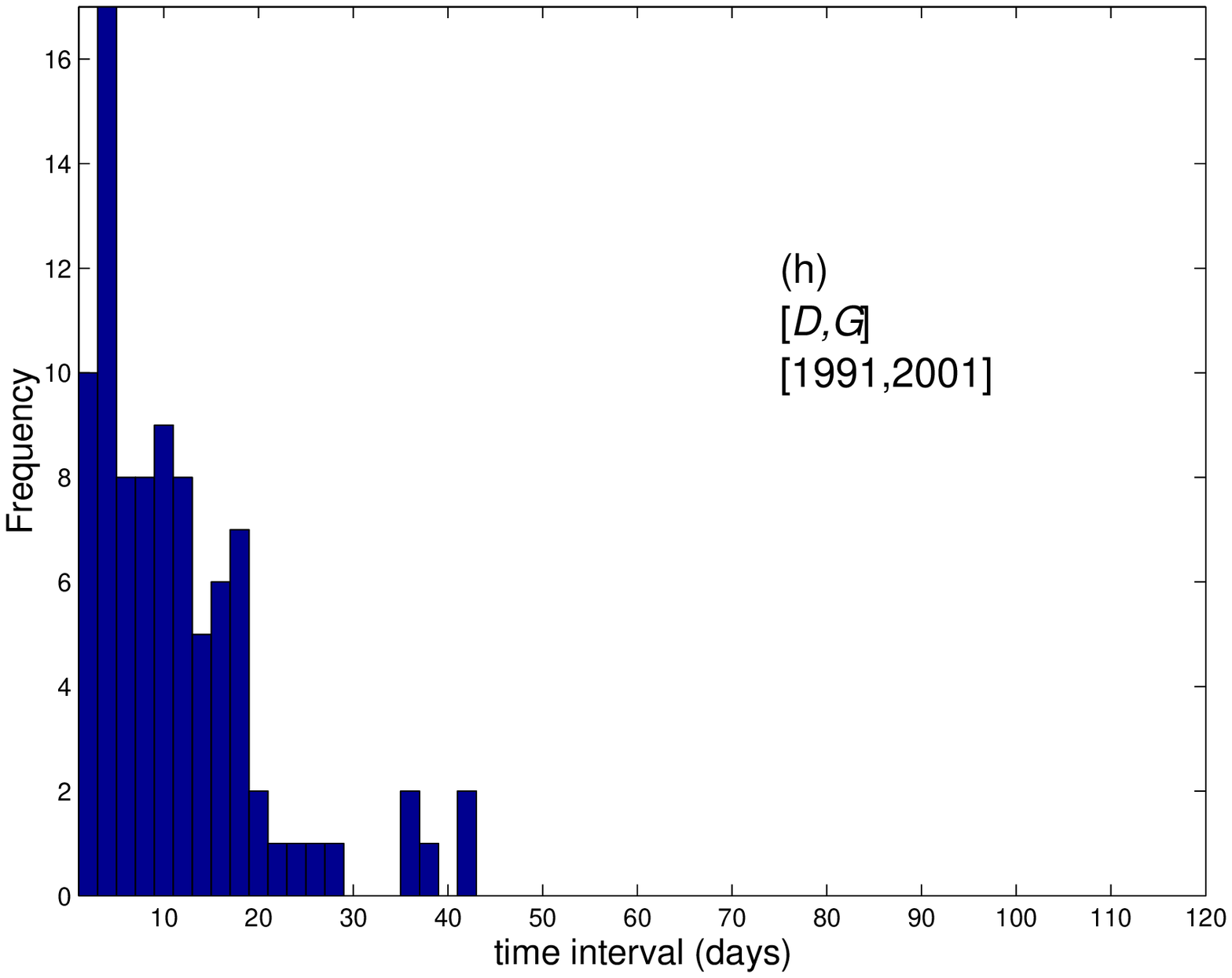}
\end{center} \caption{Frequency of time interval between consecutive
$D$ and $G$ crosses in different combinations, e.g. $GG,DD, 
GD,DG$ for the
S\&P500 closing price signal. Two periods are considered: (a-d) from Jan 1, 
1980 to Dec
31, 1990; (e-h) from Jan 1, 1991 to Dec 31, 2001.} \end{figure}

\newpage \begin{figure}[ht] \begin{center} \leavevmode \epsfysize=3.5cm 
\epsffile{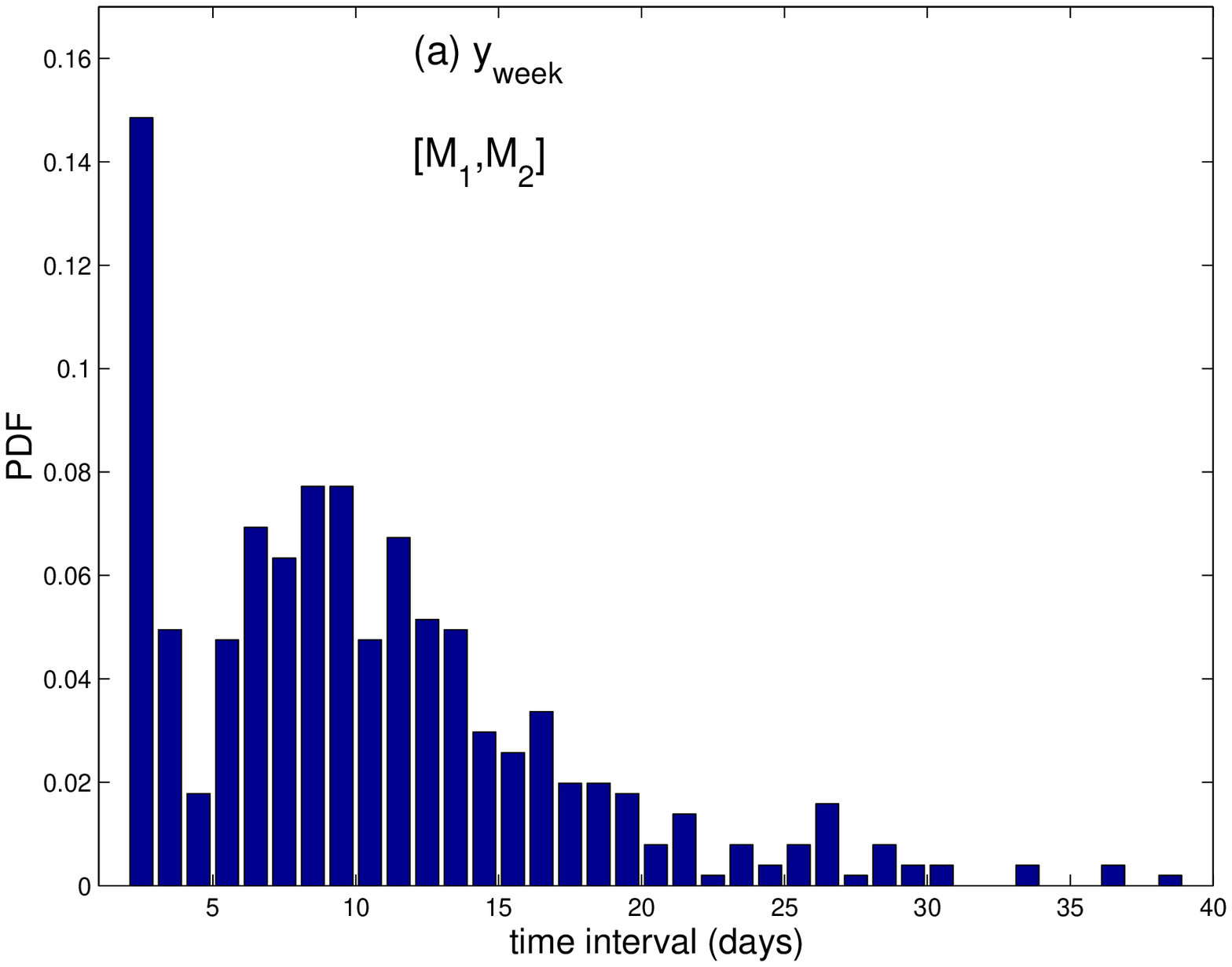}
\hfill \leavevmode \epsfysize=3.5cm \epsffile{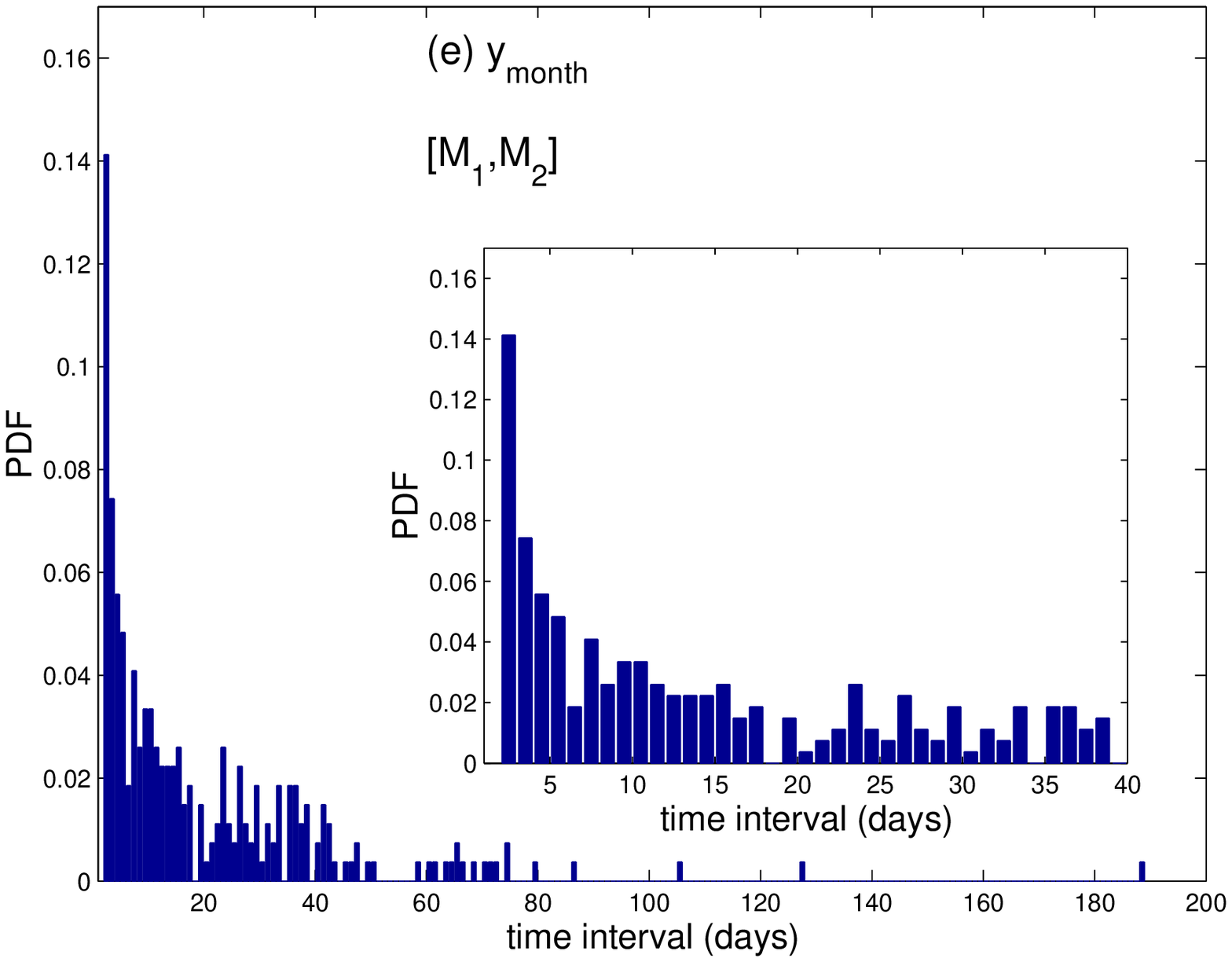}
\vfill \leavevmode \epsfysize=3.5cm \epsffile{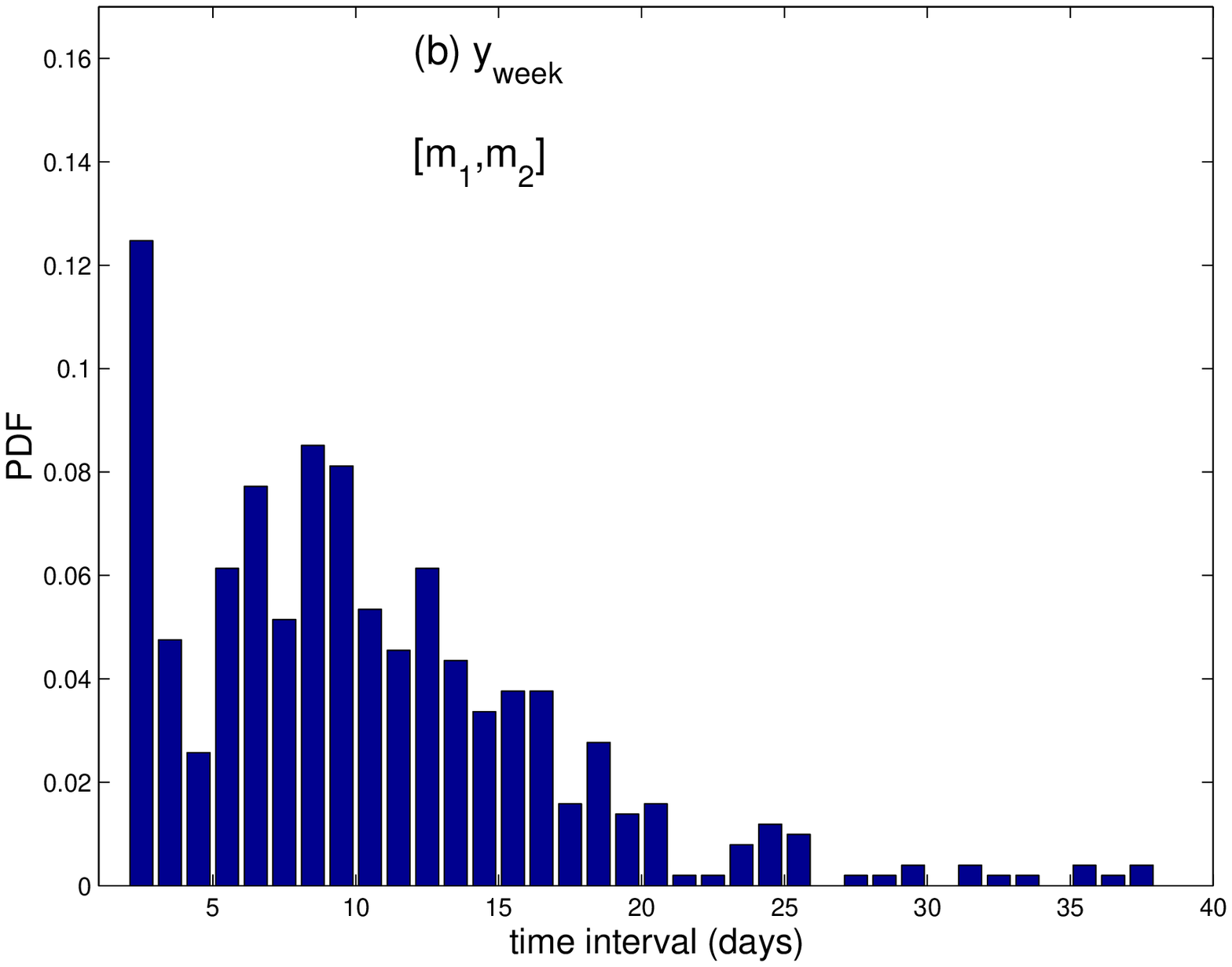}
\hfill \leavevmode \epsfysize=3.5cm \epsffile{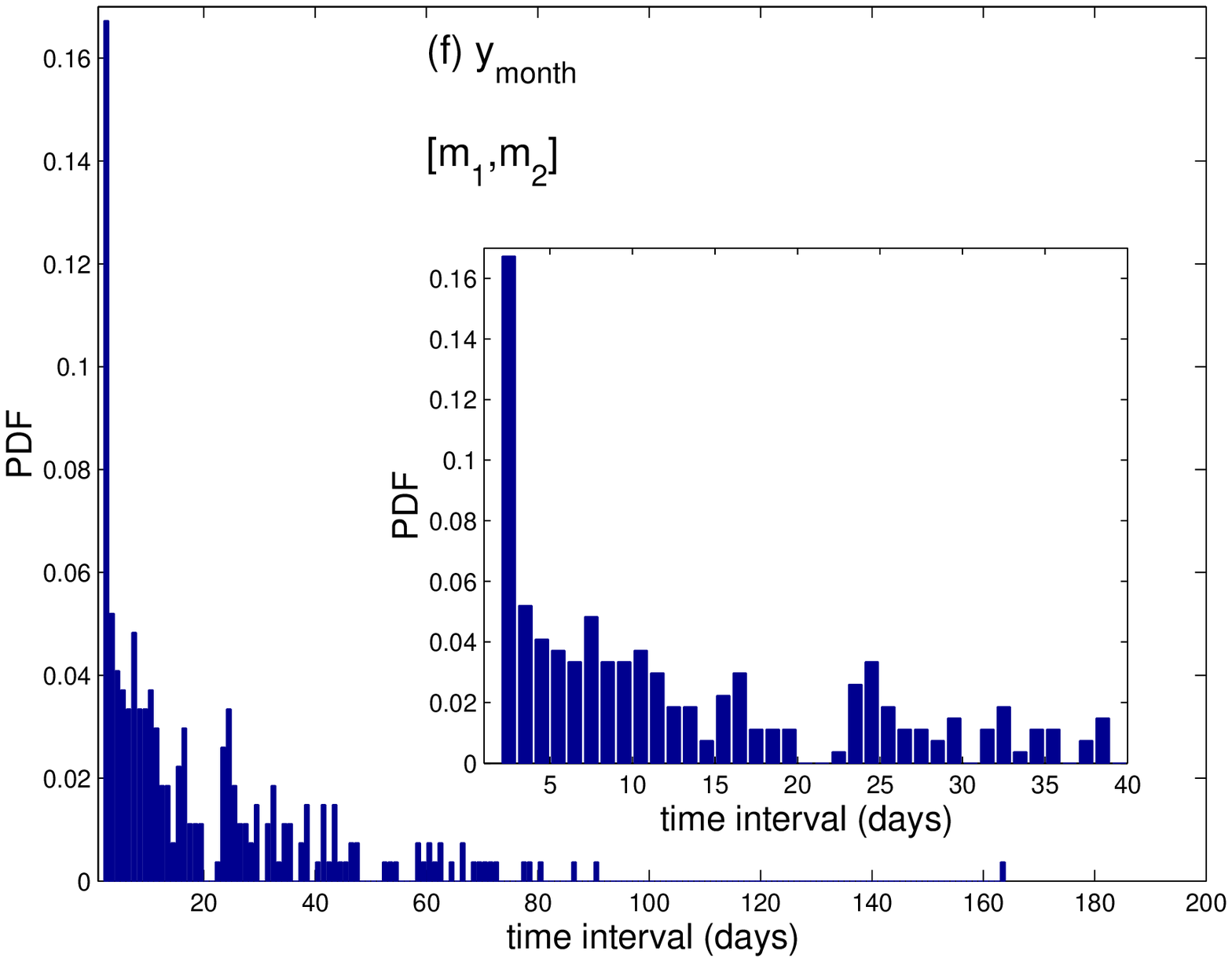}
\vfill \leavevmode \epsfysize=3.5cm \epsffile{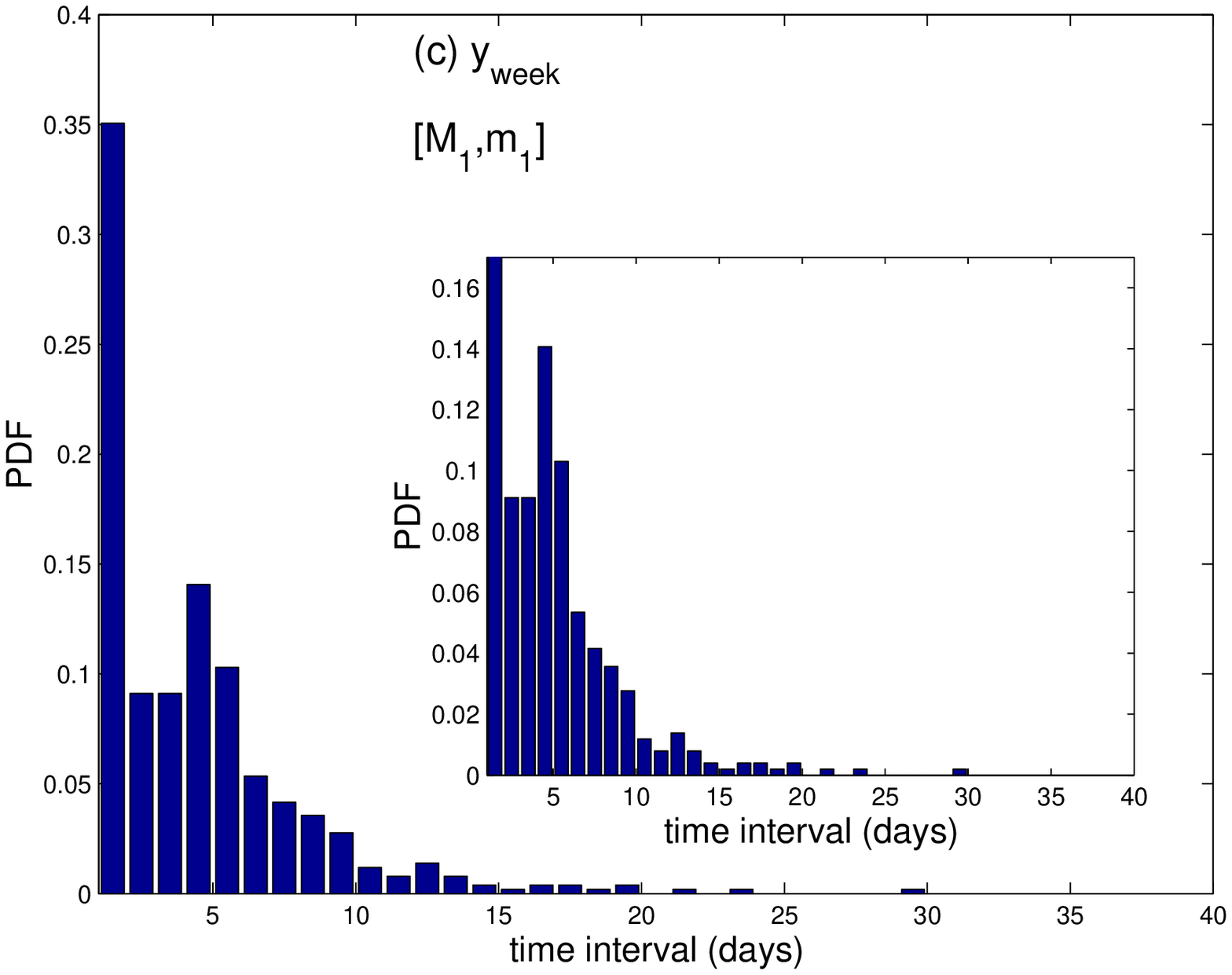}
\hfill \leavevmode \epsfysize=3.5cm \epsffile{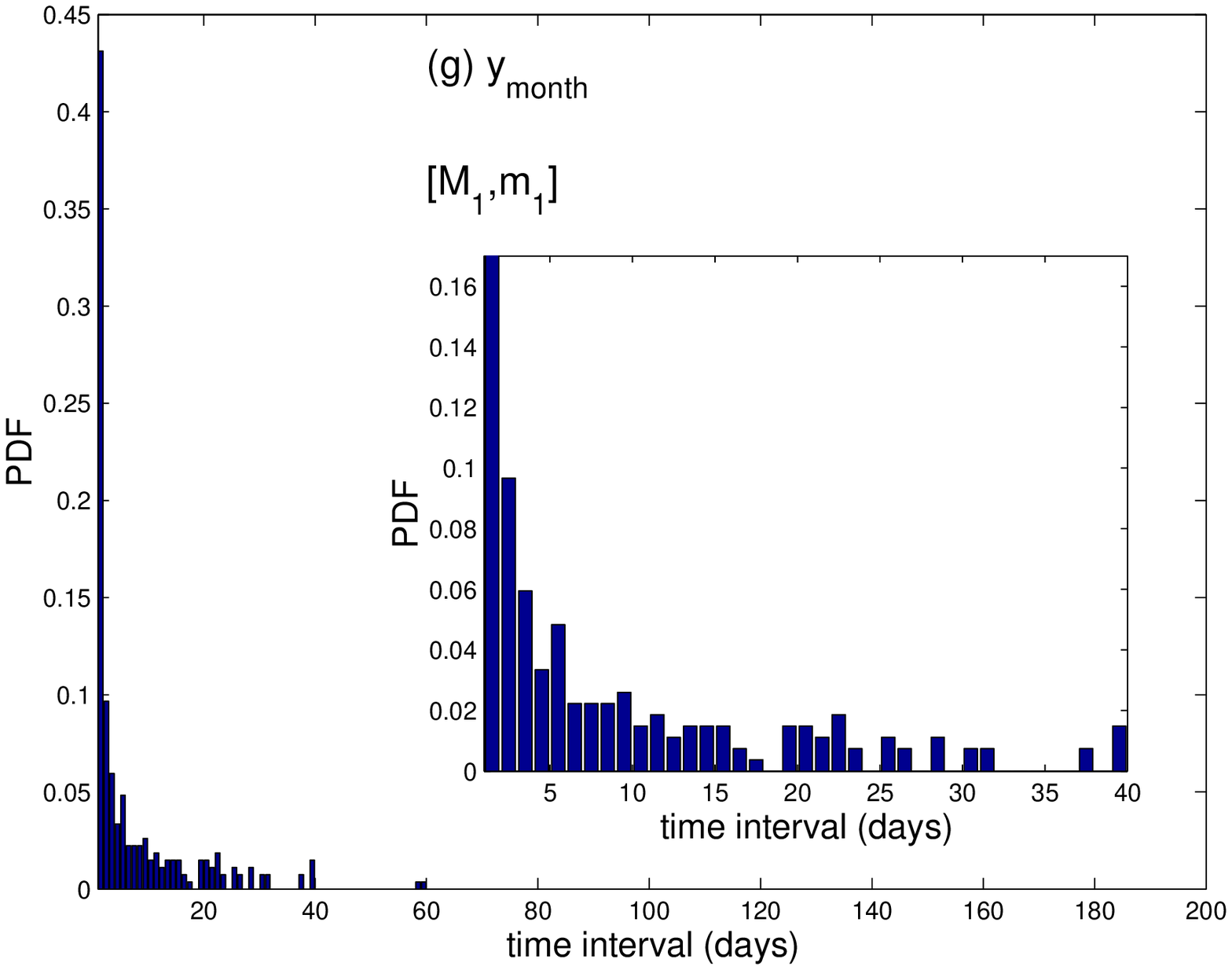}
\vfill \leavevmode \epsfysize=3.5cm \epsffile{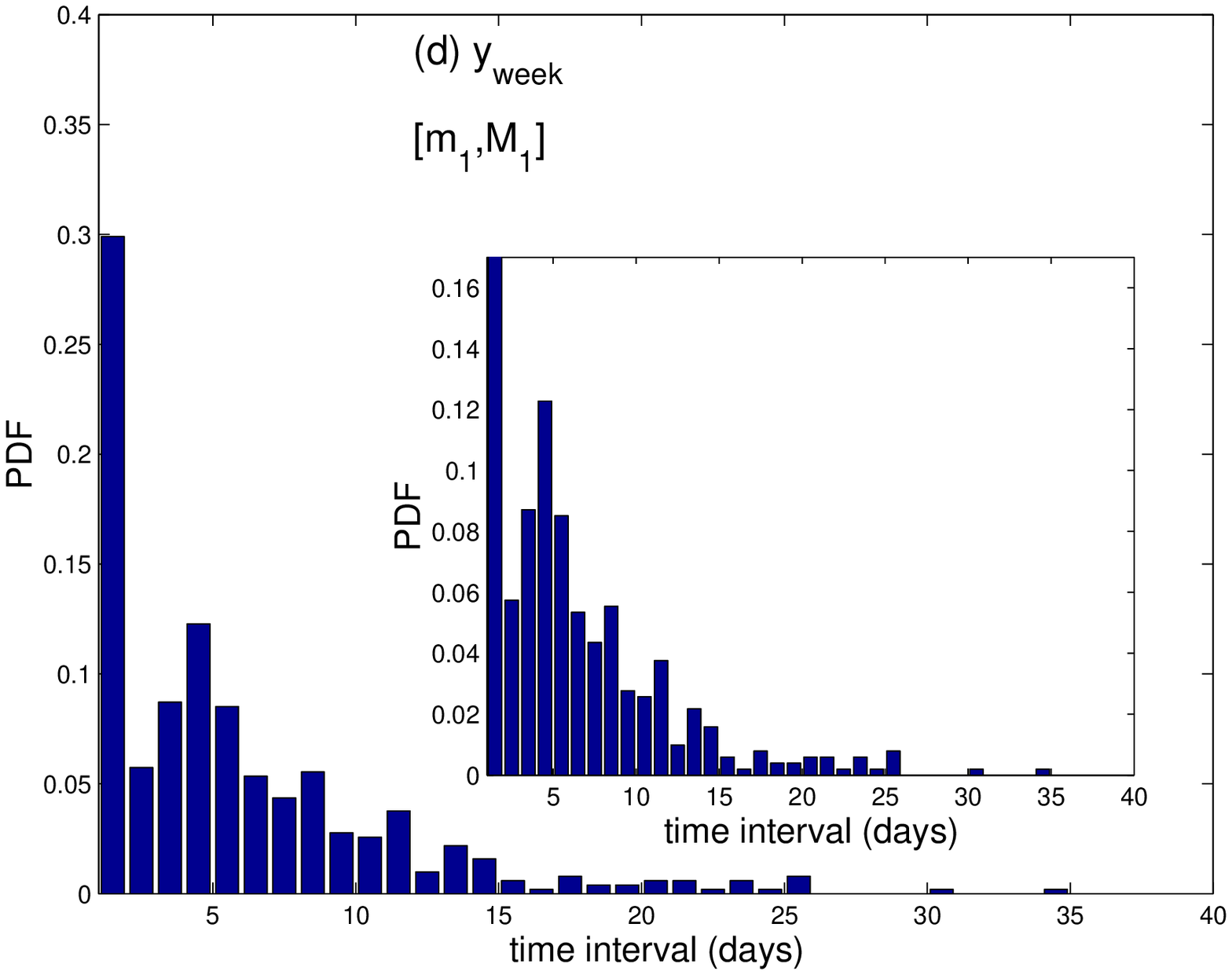}
\hfill \leavevmode \epsfysize=3.5cm \epsffile{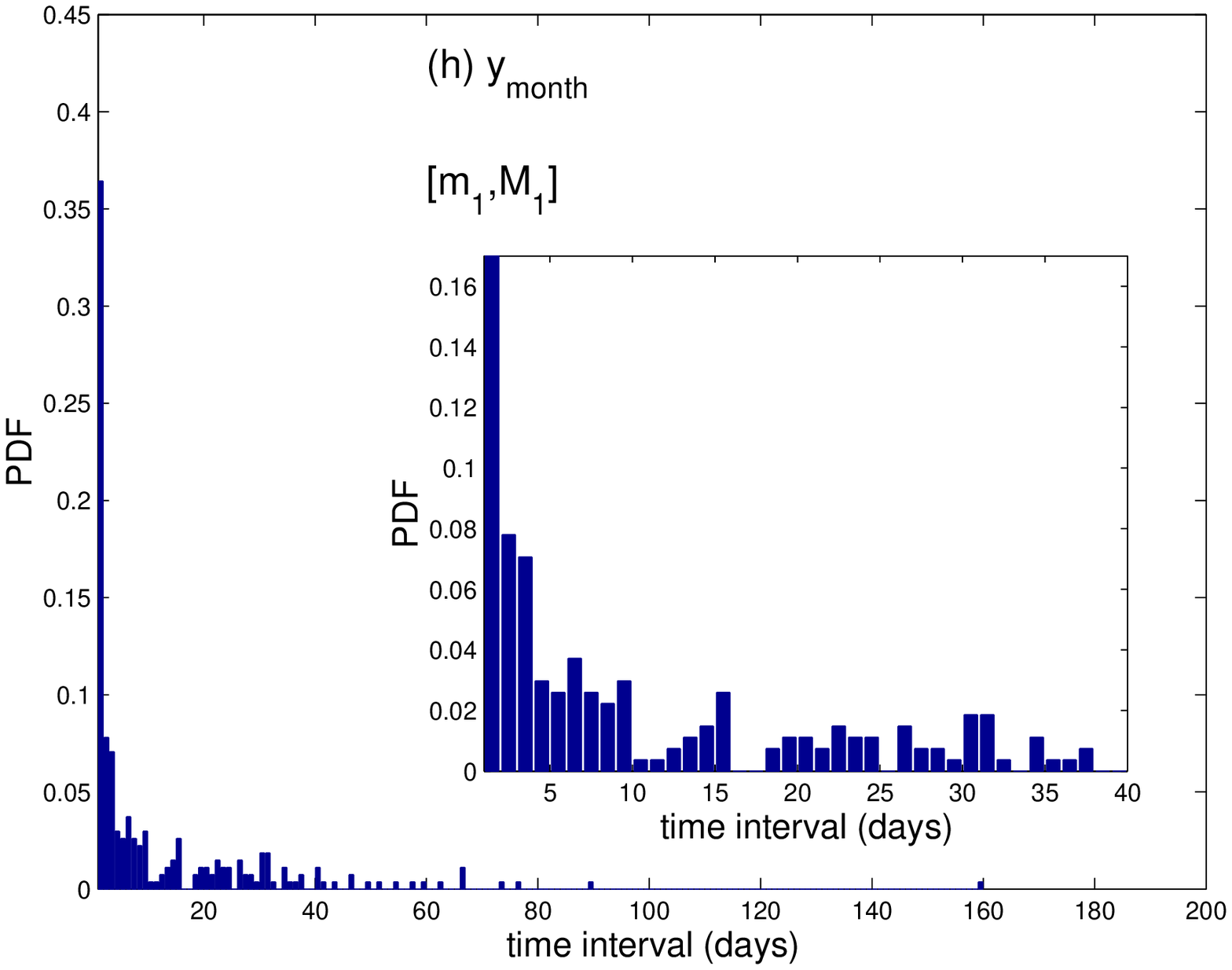}
\end{center} \caption{Probability density function of time interval
between consecutive  (a) $maxima$ and (b) $minima$ for 5-day (weekly) moving
average $y_{week}$, and (c) $maxima$ and (d) $minima$ for 21-day (monthly)
moving average $y_{month}$ of the S\&P500 closing price signal for the period
from Jan 1, 1980 to Dec 31, 2001.} \end{figure}

\newpage \begin{figure}[ht] \begin{center} 
\leavevmode \epsfysize=4cm
\epsffile{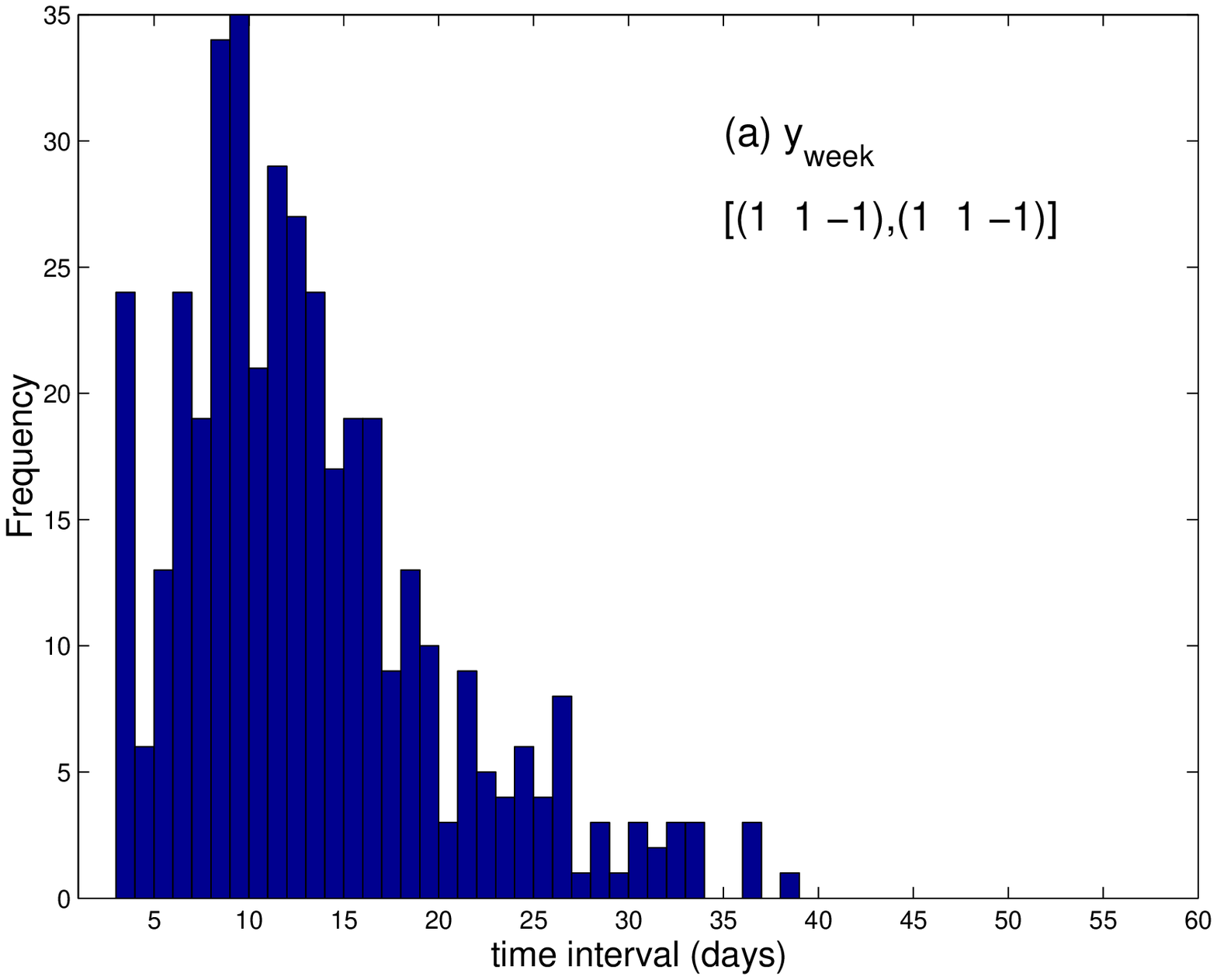} 
\hfill \leavevmode \epsfysize=4cm \epsffile{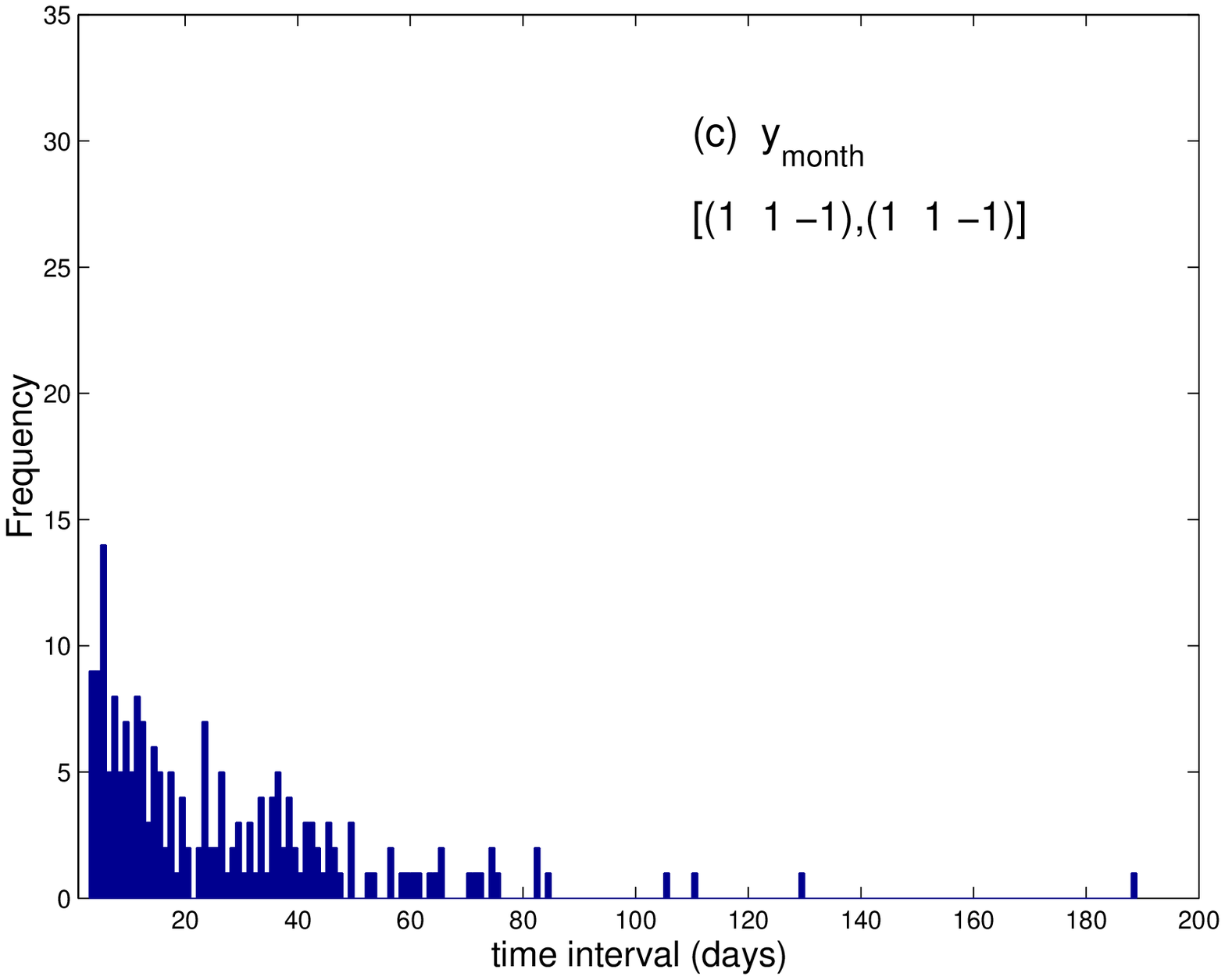}
\vfill \leavevmode \epsfysize=4cm \epsffile{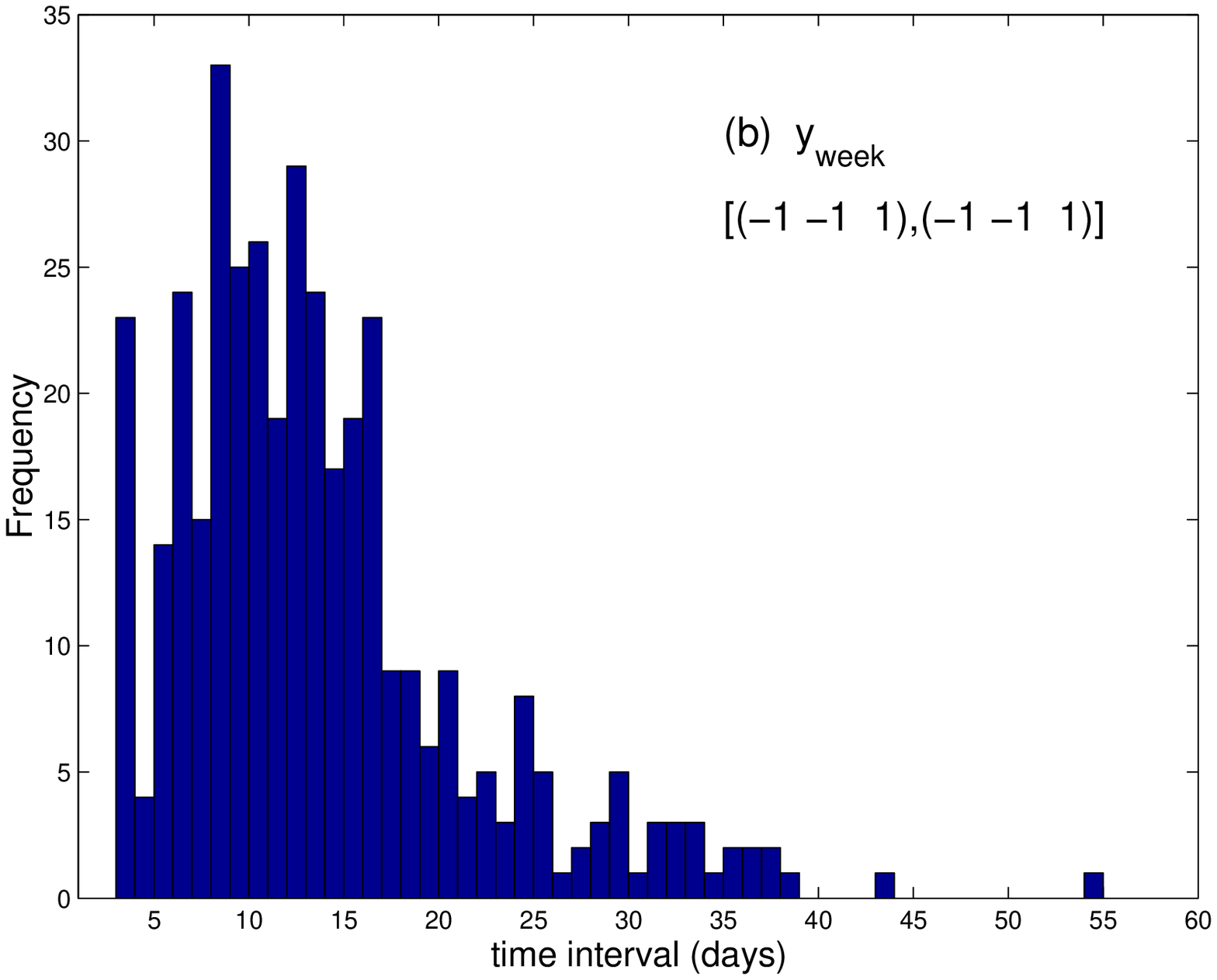}
\hfill \leavevmode \epsfysize=4cm \epsffile{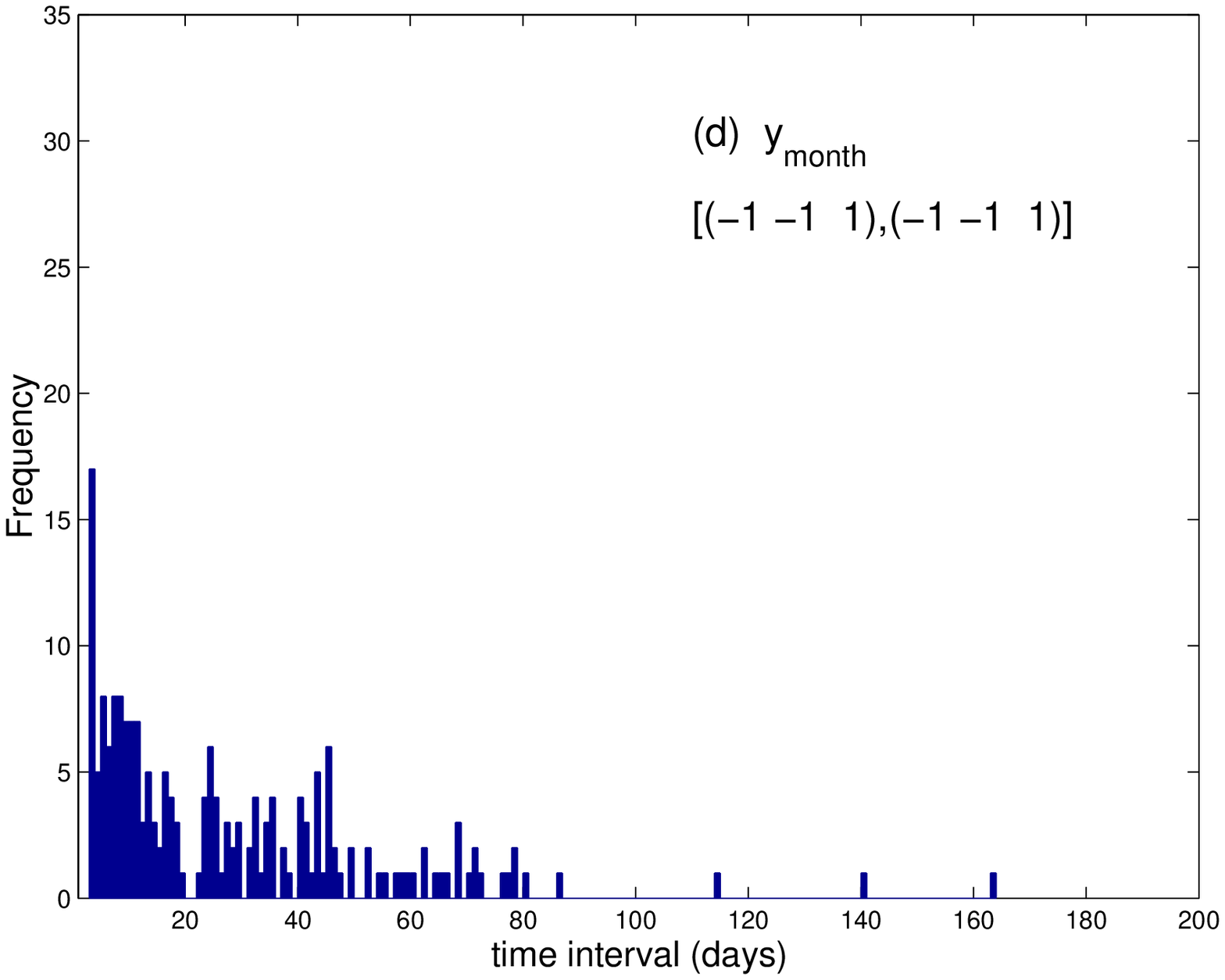}
\end{center} \caption{Frequency of time interval appearance between
consecutive (a) (1 1 -1) and (b) (-1 -1 1) for 5-day (weekly) moving average
$y_{week}$, and (c) (1 1 -1) and (d) (-1 -1 1) for 21-day (monthly) moving
average $y_{month}$ of the S\&P500 closing price signal for the 
period from Jan
1, 1980 to Dec 31, 2001.
} \end{figure}


\begin{thebibliography}{99}
\bibitem{footnote} Strictly speaking, in accord with the notation established
in Eqn. (1), these should be labelled $y_{T_{1}}$ and $y_{T_{2}}$, but we
will prefer the lighter notation established in the main text whenever no 
confusion can arise.
\bibitem{nomenclature} The vivid terms ``death" cross and ``gold" cross were 
introduced in the physics literature by Vandewalle and Ausloos.\cite{nvmav2}
They are commonly used in the French financial literature (``croix de
mort" and ``croix dor\'ee"), although they are not as widespread in 
anglophone financial sources.
\end{thebibliography}
\end{document}